\documentclass[aps, twocolumn,superscriptaddress, longbibliography]{revtex4-2}
\usepackage{lipsum,adjustbox}
\usepackage{bm}
\usepackage{float}
\usepackage{lipsum}
\usepackage{braket}
\usepackage[mode=buildnew]{standalone}
\usepackage{amsmath}
\usepackage{amssymb}
\usepackage{physics}
\usepackage[english]{babel}
\usepackage{siunitx}
\usepackage{graphicx}
\usepackage[normalem]{ulem}
\usepackage{colortbl}
\usepackage{xcolor}
\usepackage{soul}
\usepackage{tikz}
\usetikzlibrary{quantikz}
\usepackage[utf8]{inputenc}
\usepackage{pgfplots}
\usetikzlibrary{matrix}
\pgfplotsset{compat=newest}
\usepackage{pgfkeys}
\usepackage{algorithm}
\usepackage{algpseudocode}
\usepackage{setspace}
\usepackage[thinlines]{easytable}

\newcommand*\MyScale{1}

\tikzset{%
	every picture/.style={%
		scale=\MyScale,
	}
}

\newcommand{\red}[1]{{\color{black}#1}}

\DeclareUnicodeCharacter{2212}{−}
\usepgfplotslibrary{groupplots,dateplot}
\usetikzlibrary{patterns,shapes.arrows}
\pgfplotsset{compat=newest, every axis legend/.append style={ at={(1.05,0.95)}, anchor=north west,legend columns = 1}}
\usetikzlibrary{shapes,positioning}

\usepgfplotslibrary{groupplots}

\pgfplotsset{every x tick label/.append style={font=\large, yshift=0.5ex}}
\pgfplotsset{every y tick label/.append style={font=\normalsize, xshift=0.5ex}}
\pgfplotsset{minor tick style={draw=none}}

\usepackage[hidelinks,colorlinks=true,linkcolor=blue,citecolor=blue]{hyperref}
\usepackage{cleveref}

\begin{document}
	\title{Continuous quantum gate sets and pulse class meta-optimization}
	\author{Francesco Preti}
	\email{f.preti@fz-juelich.de}
	\affiliation{Forschungszentrum J\"ulich, Institute of Quantum Control (PGI-8), D-52425 J\"ulich, Germany}
	\affiliation{Institute for Theoretical Physics, University of Cologne, D-50937 K\"oln, Germany}
	\author{Tommaso Calarco}
	\affiliation{Forschungszentrum J\"ulich, Institute of Quantum Control (PGI-8), D-52425 J\"ulich, Germany}
	\affiliation{Institute for Theoretical Physics, University of Cologne, D-50937 K\"oln, Germany}
	\author{Felix Motzoi}
	\email{f.motzoi@fz-juelich.de}
	\affiliation{Forschungszentrum J\"ulich, Institute of Quantum Control (PGI-8), D-52425 J\"ulich, Germany}

	\date{\today}
	
	\begin{abstract}
    Reduction of the circuit depth of quantum circuits is a crucial bottleneck to enabling quantum technology. This depth is inversely proportional to the number of available quantum gates that have been
    synthesized. Moreover, quantum gate-synthesis and control problems exhibit a vast range of external
    parameter dependencies, both physical and application specific. In this paper, we address the possibility of learning families of optimal-control pulses that depend adaptively on various parameters, in order
    to obtain a global optimal mapping from the space of potential parameter values to the control space
    and hence to produce continuous classes of gates. Our proposed method is tested on different experimentally relevant quantum gates and proves capable of producing high-fidelity pulses even in the presence of multiple variables or uncertain parameters with wide ranges.
	\end{abstract}
	\maketitle
	
	\section*{Introduction}
	The standard view of quantum computation \cite{NielsenChuang2010} uses the classical-computing abstraction of a subdivision into a finite set of gates, measurements, and state reset tasks. This paradigm has a number of benefits:~notably, it permits formal derivation of universal computation \cite{dawson2005solovay,Barenco1995}, that is, the composition of a quantum circuit for any desired unitary operation, as well as error-correcting codes \cite{Asher1985, gottesman1998theory}, where specific error syndromes can be measured and corrected.  On the other hand, this abstraction abandons the essential analog character of quantum devices, from which they have the most to gain or lose in terms of their expressibility or fragility, respectively. That is, the power of quantum processors depends strongly on the number of usable gates available to them.
	
	Practically speaking, the usage of discrete gate sets falls short in at least four important respects.  First, the analog character is a more complete (and therefore efficient) description of variability between different qubits, which is inevitable, for instance, in solid-state qubits \cite{arute_quantum_2019}. The use of qubit-agnostic gate sets as the computational primitive means that each qubit must be individually optimized and calibrated to yield each such gate \cite{arute_quantum_2019}, where parametric description of the gates would naturally capture the variations. 
	Second, these variations can occur for a given qubit as a function of time \cite{Proctor2020}, e.g., due to time-dependent noise, and require complete recalibrations where, typically, drift involves only a single parameter. This is very taxing on the routines for error suppression, mitigation, and correction. 
	
	Third, the complexity arising from parameter variations in space and time is exacerbated by the subsequent circuit complexity in composing useful circuits.  It is well known that, while discrete gate sets can be universal, the required number of discrete  constituent gates can be very large even for a simple circuit \cite{Barenco1995, foxen2020demonstrating, yao2021reinforcement}. \red{Allowing for analog parameter tuning can dramatically increase the controllability of the system and thereby necessarily reduce the depth of the quantum circuit for arbitrary tasks (see, e.g., Appendix~\ref{appx:compilation}).} Since errors accrue with circuit depth (notably due to decoherence), this increase in the circuit success probability may be highly beneficial to both short-term and long-term (i.e., fault-tolerant) approaches. 
	
	Fourth, a further optimization layer is currently ubiquitous in noisy intermediate-state quantum (NISQ) algorithms \cite{preskill2018quantum}. Such optimizable circuits include adiabatic \cite{Albash2018, Barends2016}, annealing \cite{Apolloni1989, Kadowaki1998}, and variational \cite{peruzzo2014variational, farhi2014quantum} quantum algorithms. The common denominator in these approaches is the circuit being treated as a black box, with a set of analog parameters acting as knobs to tune as inputs for the respective algorithm. These analog inputs act as terms in the Hamiltonian and thus may generate various quantum continuous gate sets. These cannot realistically be compiled with digital gates, and, moreover, to ensure that the gate set is correctly specified requires a formal approach for their general construction. 
	
	\begin{figure*}
		\hspace{4.0cm}
		\includegraphics[width=\textwidth]{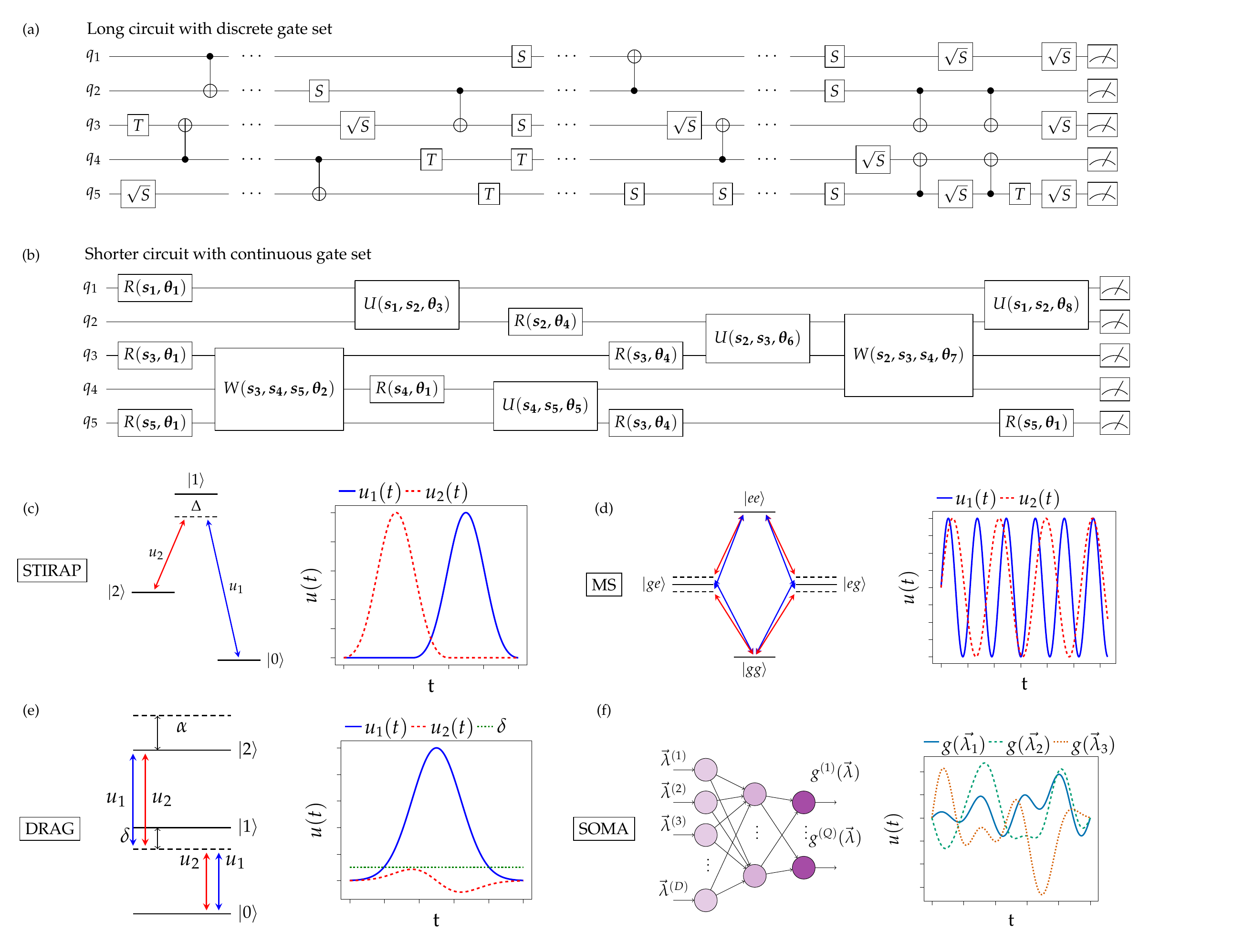}
		\hspace{-1cm}
		\caption{An explanatory diagram of the concepts discussed in Section \ref{sec:cgl}. (a) A general quantum circuit containing a long sequence of discrete unitaries (Clifford + $T$ set), which do not exhibit any dependence on continuous parameters. (b) A general quantum circuit containing different unitaries with different continuous parameters $s_1, s_2, ..., s_5$ for different qubits $q_1, q_2, ..., q_5$ and angles $\theta_1, \theta_2 ,..., \theta_8$ parametrizing each gate. $R$, $U$, and $W$ represent an analog single-, two- and three-qubit gate, respectively. Here, the same unitary parametrization is capable of representing all the different gates needed in the circuit. We show some notable analytical solutions used to engineer specific gate operations, which are usually implemented due to their adaptive character and simplicity (see Eqs.~\ref{eq:analytic_solutions},\ref{eq:ms}, and \ref{eq:drag}). (c) STIRAP \cite{kuklinski1989adiabatic,vitanov2017stimulated}. (d) The M{\o}lmer-S{\o}rensen gate \cite{Soerensen1999_1, sorensen2000entanglement}. (e) DRAG \cite{Motzoi2009, theis2018counteracting}. (f) The method that we propose, SOMA, does not make strict assumptions on how the pulse depends on the problem parameters, but, rather, discovers it more generally through training.}\label{fig:SOMA}
	\end{figure*}
	
	Our contribution, in this work, is to present a unified framework to efficiently describe and optimize continuous quantum gate sets in these scenarios.  This framework allows learning of the parametric gates that can be tuned to very high fidelity across a large number and wide range of parameter values.  We refer to our method as single-optimization multiple-application (SOMA)  quantum gate synthesis.  This kind of learning can be understood as an instance of meta-optimization \cite{hospedales2020metalearning,Devikanniga2019} or adaptive-trajectory learning \cite{Arimoto1984, Tang2018, Souza2001, Jetchev2009, Chen2019, Ozaki1991, Boettcher2022} \red{and can be related to more recent uses of neural networks in quantum physics \cite{Bukov2018, Yang2018, Niu2019, Wu2019, Lohani2020, yao2021reinforcement}.} That is, we find a solution for an optimizer that itself provides solutions to specific problem instances or, with a different formulation, automatically discovers heuristics to construct solutions for a specific optimization problem. In particular, we see that our algorithms are able to synthesize heuristics for general Hamiltonians.
	
	We break down the problem of generalized gate synthesis into the following components.  We parametrize our quantum gate set using continuous indices that represent either physical system parameters or desired angles of a continuous Lie group. We then present two different machine-learning methods for obtaining continuous control parametrizations that generate the indexed gate set. The first is a supervised approach where traditional optimal-control theory is used to generate an operational data set from which a generalized gate-set recipe can be trained. The second is an unsupervised approach using back propagation from which the continuous gate set can be incrementally learned over the entire training population.  Finally, we show that our approaches encompass the various situations discussed above, including general solutions for gates given generic physical architectures with wide parameter ranges, noise-adaptive optimal-control theory, and compilation of a Lie group instead of a single element. 
	
	The paper is organized as follows. In Sec.~\ref{sec:cgl}, we introduce the notation for supervised and unsupervised training of parametrized pulses. In Sec.~\ref{sec:results}, we discuss the results obtained by applying these methods to single-qubit and two-qubit gates in the presence of leakage, showing that they display similarities with known analytical-solution families. Furthermore, we also investigate how our methods perform compared to other existing algorithms. \red{Finally, in Sec.~\ref{sec:robvsada} we analyze the dependence on the parameter variability range, the training data set size and batch size, the system and network size and again compare our algorithms to other numerical robust approaches.} We summarize our conclusions in Sec.~\ref{sec:conc}.

	\section{Continuous gate set learning} \label{sec:cgl}
	\subsection{Definitions}
	We define a continuous gate set in terms of some continuous sets or distributions of $n$ system parameters $s_1, s_2, \dots, s_n$ and $m$ gate specifications $\theta_1, \theta_2, \dots, \theta_m$. An element of such a gate set is a unitary transformation between two Hilbert spaces $H_1$ and $H_2$:
	\begin{align}\label{eq:gate set}
	U(s_1, s_2, \dots, s_n,\theta_1, \theta_2, \dots, \theta_m): H_1 \mapsto H_2.
	\end{align}\\
	Obtaining a single element of this set is a well-known problem in quantum information. Depending on whether the unitary is synthesized from discrete or analog dynamics, its composition is referred to as circuit compilation \cite{Barenco1995, dawson2005solovay, Bocharov2015,Motzoi2017} or optimal-control theory \cite{Warren1993Dream, freeman1998spin, Khaneja2005, goerz2019krotov, Caneva2011, GlaserCat2015}, respectively. 
	
	In Fig.~\ref{fig:SOMA}(a), we see an example of a generic circuit acting on different qubits. Each qubit in space (and time) will have different values of the common system parameters $\{s_k\}$. In addition, the different unitaries in the gate set $\{U_i\}$, each additionally characterized by rotation angles $\{\theta_{i,j}\}$, may vary both throughout the circuit and in iterated uses of the circuit (e.g., in NISQ algorithms). For compactness we now regroup the continuous indices into a common array of indices $\vec \lambda = (s_1, s_2, \dots, s_n,\theta_1, \theta_2, \dots, \theta_m)$.
	
	Such a generalization of the gate-synthesis problem can be framed as solving for the inverse function of the general dynamics given in Eq.~\eqref{eq:gate set}, that is, for
	\begin{align}
	g: \vec \lambda \mapsto \bm u(t),
	\end{align}
	where $\bm u(t)$ is a wave-form function in the standard case of optimal-control theory, but can also be thought of as a discrete sequence of hard pulses, as in NMR applications, or of unitary gates in circuit compilation. Importantly, the optimal $\bm u(t)$ changes for each parametrization $\vec\lambda$ of the unitary, which can be generated from e.g.~the Schr{\"o}dinger equation. 
	\begin{align}
	\dot U(\vec{\lambda}, t) = -i H (\vec \lambda, \bm u(t)) U(\vec{\lambda}, \bm{u}(t)),
	\end{align}
	or other equations of motion defining the system. This task can be cast as an instance of meta-optimization  \cite{Grefenstette1986} or trajectory learning for control \cite{Schaal2010, Atkeson1986, Ozaki1991}.
	We formalize the 
	meta-optimization problem as list of problem-parameter vectors $\vec{ \lambda}_1, ..., \vec{\lambda}_L \in V_{\lambda} \subset \mathbb{R}^D $ with figures of merit $F_1(\bm{x}) = F(\bm{x}, \vec{\lambda}_1), ..., F_L (\bm{x})= F(\bm{x}, \vec{\lambda}_L)$ and initial guess $\bm{x}_0$, the physical parameters of which vary somewhat from each other, such that $\vec{\lambda}_i \sim \pi(\vec{\lambda} \vert \bm{v})$ for $0 \leq i \leq L$ and $\bm{v} \in V_{\lambda}$,  is drawn from a parameter distribution. Throughout the paper we assume, without loss of generality, $\pi$ to be a multidimensional uniform distribution, such that $\pi(\vec{\lambda} \vert \bm{v}) = U(\vec{\lambda}_{\text{min}},\vec{\lambda}_{\text{max}})$, with $\bm{v} = (\vec{\lambda}_{\text{min}},\vec{\lambda}_{\text{max}})$ defining the parameter space $V_{\lambda}$.

    The objective is to find optimal parameters $\bm{w}^*$,
	\begin{align}
	\bm{w}^* = \underset{\bm{w}}{\text{argmin}} \ \{1 - F_i(\bm{w}) \vert \ \forall i=1,...,L \}
	\end{align}
	which allow for simultaneous optimization of all the systems considered within the range of sampled parameters.
	
	\subsection{Analytical adaptive control}\label{sec:ansol}
	
	The most straightforward way to generate classes of solutions to quantum gates has been the development of analytical solutions for particular quantum systems.
	
	They have in common the knowledge of the relevant state of the system at all times during the evolution. In particular, analytical knowledge of the eigenvalues \cite{li2021non} allows reverse engineering of the pulses to provide (near) exact solutions for the desired states.
	
	Fig.~\ref{fig:SOMA}(c)-(f) shows several celebrated examples where such general classes of solutions have been found. We quickly review some of their main features. Given a trial pulse shape such as a Gaussian envelope $p(t,t_1,t_2,\theta)=\mathcal{A} e^{( t-t_2/2+t_1/2)^2/\sigma^2}$, where $\theta$ denotes the area under the curve, the following dynamical solutions have been found.
	
    Fig.~\ref{fig:SOMA}(c) shows the stimulated Raman adiabatic passage (STIRAP) solution for transfering population between disconnected states $\ket{0}$ and $\ket{2}$, while avoiding any (nonvanishing) temporary population in the intermediary connecting state $\ket{1}$. The pulse shaping is given by
	\begin{align}\label{eq:analytic_solutions}
	&\text{STIRAP}(\ket{0}\rightarrow\ket{2}): (\Omega_1, \Omega_2, \Delta, \theta,T)\mapsto \bm u(t),\\
	&\quad\quad  u_1(t) = \Omega_1 e^{i\theta} p(t,T/3,T,\pi)e^{i\Delta t} \ketbra{0}{1}\nonumber  \\
	&\quad\quad u_2(t) = \Omega_2 p(t,0,2T/3,\pi)e^{-i\Delta t} \ket{1}\bra{2} \nonumber
	\end{align}
	where the order of the pulses $u_1$ and $u_2$ is famously counterintuitive \cite{kuklinski1989adiabatic,vitanov2017stimulated}.
	
	Fig.~\ref{fig:SOMA}(d) shows the M{\o}lmer-S{\o}rensen (MS) method for trapped-ion gates \cite{Soerensen1999_1, sorensen2000entanglement}. The required laser pulses are given by
	\begin{align}\label{eq:ms}
	&\text{MS}(\sigma_x\otimes\sigma_x\otimes\dots\otimes\sigma_x): (\Omega_1, \Omega_2, \delta, \theta_1, \theta_2)\mapsto \bm u(t),\\
	&\quad\quad  u_1(t) = \Omega_1  p(t,0,T,\theta_1) e^{i\theta_2} e^{i\delta t} \ketbra{g,n}{e,n-1} \nonumber\\
	&\quad\quad u_2(t) = \Omega_2 p(t,0,T,\theta_1) e^{-i\delta t} \ketbra{g,n}{e,n+1},\nonumber
	\end{align}
	where the first state index denotes the state of the relevant qubit and second index denotes the phonon occupation number. Because of the symmetry of the gate, it can in theory be used on an arbitrarily large number of qubits, i.e., it is a collective gate \cite{martinez2016compiling, Motzoi2017}.
	
	Fig.~\ref{fig:SOMA}(e) shows the Derivative Removal for Adiabatic Gate (DRAG) solution, for leakage suppression in multi-level systems \cite{Motzoi2009,theis2018counteracting}. This pulse profile is given by
	\begin{align}\label{eq:drag}
	&\text{DRAG} (\ket{0}\leftrightarrow\ket{1}): (\Omega_1, \Omega_2, \alpha, \delta, \theta_1, \theta_2)\mapsto \bm u(t), \\
	&\quad\quad  u_1(t) =  e^{i\theta_2} e^{i\delta t} p(t,0,T,\theta_1) (\Omega_1\ketbra{0}{1}+\Omega_2\ketbra{1}{2})\nonumber\\
	&\quad\quad u_2(t) = ie^{i\theta_2}e^{i\delta t}\partial_t p(t,0,T,\theta_1)/\alpha (\Omega_1\ketbra{0}{1}+\Omega_2\ketbra{1}{2}).\nonumber
	\end{align}
	
	We see, with these families of solutions, the common trend that they allow for different known system parameters or for different rotation or phase angles. Naturally, this is just a representative set, but where the equations of motion are integrable such solutions are numerous in the literature.
	
	However, there are a few evident concerns about finding such analytical solutions. Firstly, it is a labour-intensive task with no guaranteed result, where even particular solutions do not preclude that a more systematic search would produce better results. Secondly, it is important to emphasize that these are solutions to idealized models, and in practice the more accurate physical models are not exactly solved by these ansätze. Finally, most physical systems have to date not been able to find general analytical solutions beyond qubits, qutrits and highly symmetric systems, both because of the larger state space and the larger parameter space. This limits their viability for quantum computing, which requires much larger Hilbert spaces.
	
		\begin{figure*}[ht!]
		\includegraphics[width=\textwidth]{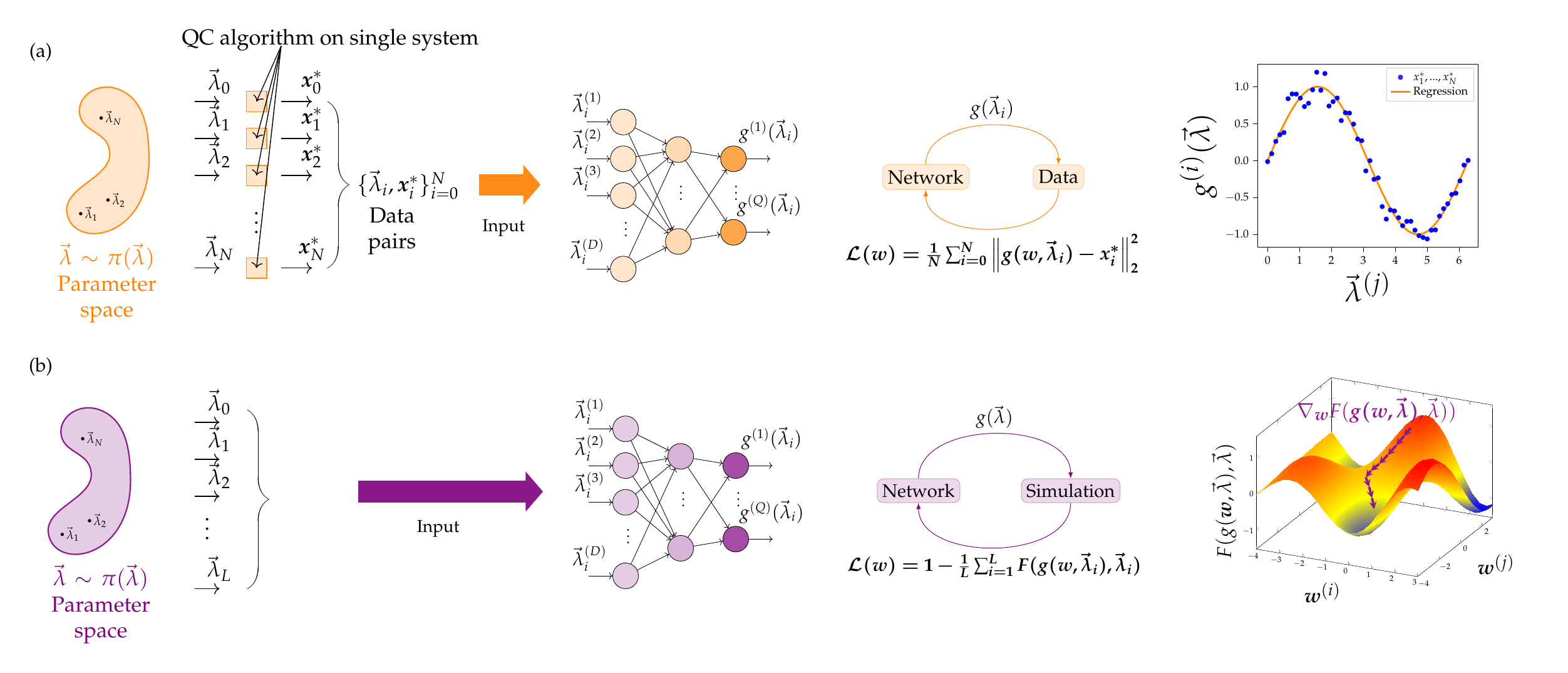}
		\caption{The Single-Optimization Multiple-Application gate synthesis method represented in its two variants. (a) In SOMA SL we first optimize $N$ different QOC problems with different problem parameters $\vec{\lambda}_1, \vec{\lambda}_2, ..., \vec{\lambda}_N$ using an optimal quantum control algorithm such as \cite{Khaneja2005,goerz2019krotov,frank2017autonomous,dalgaard2020hessian} to obtain optima $\bm{x}^*_1, \bm{x}^*_2, ..., \bm{x}^*_N$, then use a function approximator to learn the mapping $g: \mathbb{R}^D \mapsto \mathbb{R}^Q, \vec{\lambda} \longrightarrow \bm{x}^*(\vec \lambda)$ between the problem parameters and the optimal pulses. (b) In SOMA BP, we sample $L$ OQC problems $\vec{\lambda}_1, \vec{\lambda}_2, ..., \vec{\lambda}_L$ and  train the function approximator to minimize the average infidelity of the ensemble of problems using back propagation, without generating optimal solutions for a single problem with a standard quantum control method.}
		\label{fig:soma_types}
	\end{figure*}
	
	\subsection{GRAPE}\label{subsec:grape}
	GRAPE (GRadient Ascent Pulse Engineering)  \cite{Khaneja2005} is a method originally developed in the context of quantum chemistry for the optimization of dynamical evolution of NMR systems. The algorithm assumes a unitary dynamics $U(t)= e^{- \int_{t_0}^t iH(\tau) d\tau}$, governed by a Hamiltonian of type:
	\begin{align}\label{eq:ham_ctrl}
	    H(t) = H_0 + \sum_{m=1}^M u_m(t) H_m
	\end{align}
	where $H_0$ is a time-independent drift Hamiltonian and $H_m$, with $m = 1,..., M$, are different suitable control Hamiltonians with corresponding control fields $u_m(t)$. In simulations, $U(t)$ is often computed through different types of Trotterization \cite{CesarRdeOliveira}.
    GRAPE provides us with an efficient gradient of the merit function with respect to the control pulse values. The merit function is usually given by the gate fidelity:
	\begin{align}\label{eq:grapeF}
	    F = \frac{1}{d^2}\vert \Tr{U G^{\dagger}} \vert^2,
	\end{align}
	where the normalization factor $d$ corresponds to the dimension of the Hilbert space and $G$ is a target unitary, which we would like to generate using the unitary dynamics. We assume a Trotter-like unitary evolution of the system of type:
	\begin{align}
	    U(T) = \prod_{j=1}^{N_{\text{evo}}} U_j(t_j, t_{j-1}),
	\end{align}
    where $U_j(t_j, t_{j-1}) = e^{-i H dt}$, with $dt = T/N_{\text{evo}} = t_{j} - t_{j-1}\ \forall j=1,..., N_{\text{evo}}$, where $N_{\text{evo}}$ defines the number of time steps used in the Trotterization. In particular, considering Eq.~\eqref{eq:ham_ctrl}, the unitary step $U_j$ reads
	\begin{align}
	    U_j = \exp{-i dt \left(H_0 + \sum_k u_k(j) H_k \right)}.
	\end{align}
	The gradient of the fidelity can be computed iteratively starting from the so-called propagated optimal state \cite{Khaneja2005}:
	\begin{align}
	    O_j = U_{j+1}...U_{N_{evo}} G U_1 ... U_{j-1},
	\end{align}
	so that the gradient approximately results in: 
	\begin{align}
	    \pdv{F}{u_k(j)} \approx \frac{2}{d^2} \Re{\Tr{UG^{\dagger}} \Tr{ i dt H_k O_j}}. 
	\end{align}
	This approach, however, cannot directly account for variations of the underlying dynamics due to e.g., stochastic noise \cite{borneman2010application}, field inhomogeneity \cite{mottonen2006}, or Hamiltonian uncertainties \cite{khani2012} and the optimal pulses output by following the native gradient direction can prove significantly worse than expected if some of the underlying problem parameters vary. A possible way around this is to switch to a robust control approach, in which the cost function \eqref{eq:grapeF} accounts for parameter shifts. A simple way \cite{Khaneja2005} is to use an average fidelity over the parameter space sampled with quasi-Monte Carlo, 
	\begin{align}
	    \bar{F} = \frac{1}{L} \sum_{l=1}^L F(\bm{w}, \vec{\lambda}_l),
	\end{align}
	where $L$ is the number of samples.
	
	\subsection{Robust control}
	
	Similar to the adaptive solutions using analytical methods, solutions for controls robust to uncertainty in parameters have been found both by analytical and numerical means \cite{freeman1998spin,mottonen2006, borneman2010application,khani2012, biercuk2009optimized, motzoi2016backaction,schirmer2020robust,muller2018noise, Yang2018}. Since this involves only a single solution and not a family thereof, this has been the preferred method for parameter variability in quantum gate sets, as they are easier to design.
	
	Robust solutions are generally defined slightly differently from the adaptive solutions, using the figure of merit
	\begin{align}\label{eq:robust}
	\bm{w}^* = \underset{\bm{w}}{\text{argmin}} \left( \ 1 - \frac{1}{L} \sum_{i=1}^L F_i(\bm{w}) \right).
	\end{align}
	This cost function is more tractable from an optimization point of view because 
	we can simply account for an ensemble of individual cost functions by taking the average of the cost functions as the objective of the optimization. 

    Robust solutions have also been found using analytic methods. In particular, a common pulse sequence for gates with robustness to amplitude noise is the Broad-Band 1 (BB1) sequence
		\begin{align}
	&\text{BB1}(\ket{0}\leftrightarrow\ket{1}): (\Omega, \Delta \Omega, \theta_1, \theta_2)\mapsto \bm u(t),\\
	&\quad\quad  u_1(t) = \Omega p(t,0,T/4,\theta_1) e^{i\phi} \ketbra{0}{1}\nonumber \\
	&\quad\quad  u_2(t) = \Omega p(t,T/4,T/2,\pi) e^{i\theta_2+i\cos^{-1}(-\theta_1/4\pi)}\ketbra{0}{1}\nonumber \\
	&\quad\quad u_3(t) = \Omega p(t,T/2,3T/4,2\pi) e^{i\theta_2+i3\cos^{-1}(-\theta_1/4\pi)}\ketbra{0}{1} \nonumber\\
	&\quad\quad  u_4(t) = \Omega p(t,3T/4,T,\pi) e^{i\theta_2+i\cos^{-1}(-\theta_1/4\pi)}\ketbra{0}{1}\nonumber 
	\end{align}
	which is independent of offsets in Rabi frequency $\Delta \Omega$ up to sixth order, $1-F=O(\Delta \Omega^6)$  \cite{wimperis1994broadband}.
	
	Likewise, when applying gates with unknown frequency offsets, the Compensation for Off-Resonance with a Pulse SEquence (CORPSE) method \cite{cummins2000use} given by
	\begin{align}
	&\text{CORPSE}(\ket{0}\leftrightarrow\ket{1}): (\Omega, \Delta, \theta_1, \theta_2)\mapsto \bm u(t)\\
	&\quad  u_1(t) = \Omega e^{i\theta_2} p\left(t,0,\frac{T}{3},2\pi+\frac{\theta_1}{2}-\sin^{-1}\left(\frac{\beta}{2}\right)\right)\ketbra{0}{1}\nonumber \\
	&\quad u_2(t) = \Omega e^{i\theta_2} p\left(t,\frac{T}{3},\frac{2T}{3},2\pi-2\sin^{-1}\left(\frac{\beta}{2}\right)\right)\ketbra{0}{1}\nonumber \\
	&\quad  u_3(t) = \Omega e^{i\theta_2} p\left(t,\frac{2T}{3},T,\frac{\theta_1}{2}-\sin^{-1}\left(\frac{\beta}{2}\right)\right)\ketbra{0}{1}\nonumber,
	\end{align}
	with $\beta = \sin(\frac{\theta}{2})$is robust to the exact value of $\Delta$, for small enough $\Delta$.
	
	Robustness has found widespread use in quantum computation where fabrication uncertainty, use of ensemble systems, and noise have made control challenging. The use of robust control is especially important where small deviations occur over time scales roughly on par with gate durations.
	Nevertheless, if deviations are not small, if they are over a much longer (or much shorter) timescale, or if very high fidelity is sought after, then typically they have limited value. This is especially the case where variability occurs as  result of design uncertainty or slow parameter drift, or when continuous gate sets are needed as for NISQ algorithms. To understand why maximum fidelities suffer as a result of improved robustness, notice that a longer pulse sequence (with multiple pulses) will necessarily incur more decoherence. Thus while pulses such as BB1 and CORPSE will reduce drift error, the overall fidelity will not be as high as a single pulse at a fraction of the duration could have yielded. We will also show this quantitatively in the next sections.
	\subsection{Supervised training method: SOMA SL}
	Rather than constructively producing such classes, or relying solely on optimal control theoretic methods, here we pursue the approach of machine learning a functional approximation to the general solutions. Function approximators are mathematical objects capable of reproducing arbitrary functions using families of functions \cite{Goodfellow-et-al-2016}. They are normally identified with neural networks and find extensive application in machine learning, data analysis, etc.
	For the supervised approach, which is sometimes referred to as trajectory learning in the robotic control literature \cite{Atkeson1986}, we employ an optimizer to find the corresponding minima for a set of problems and a regressor $g: \mathbb{R}^D \mapsto \mathbb{R}^Q$ which maps the problem parameter space to the space of solutions to the given problem.  We refer to this approach as SOMA SL (SOMA with supervised learning). A sketch of the algorithm is provided in Fig.~\ref{fig:soma_types} (a). Starting from a seed problem with solution $ \bm x^*_0$, generated previously, for $N$ different optimal quantum control problems parametrized by $\vec{\lambda}_1, ..., ,\vec{\lambda}_N$, we generate $N$ solutions $\bm{x^*}_1, ..., \bm{x^*}_N$. Then we train the neural network to find the best non-linear mapping between the original parameters and the solutions. Training is performed via a standard mean squared error (MSE) loss:
	\begin{align}\label{eq:supervisedloss}
	&\mathcal{L}(\bm{w}) = \sum_{i=1}^{N}\norm{\bm{z}^*_i - {g}(\bm{w}, \vec{\lambda}_i)}_2^2 \\
	&\bm{z}^*_i = \frac{\bm x^*_i - \bar{\bm x}}{\bm{\sigma_x}}
	\end{align}
	where $\bar{ \bm x}$ is the mean value of the generated data, $\bm{\sigma_x}$ its standard deviation, and $\bm{z}^*_i$  the normalized data.
	\begin{algorithm}[H]
		\setstretch{1.20}
		\caption{SOMA SL}
		\label{alg:SOMA1}
		\hspace*{\algorithmicindent} \textbf{Input} $\bm{w}, \vec\lambda_0, \vec\lambda_{\text{min}}, \vec \lambda_{\text{max}}$, optimizer $\text{OPT}$\\
		\hspace*{\algorithmicindent} \textbf{Output} $\bm{w}^*$
		\begin{algorithmic}[1]
			\State $\bm{x}^* = OPT(F(\bm{x},\vec{\lambda}_0), \nabla_x F(\bm{x}, \vec{\lambda}_0), \bm{x}_0)\quad\quad\quad\quad\quad\quad\quad\quad$ \Comment{with random restart}
			\For{$i=1$ to $N$}
			\State $\vec{\lambda}_i \sim \pi(\vec{\lambda}_{\text{min}}, \vec{\lambda}_{\text{max}})$
			\State $\bm{x}_i^* = OPT(F(\bm{x},\vec{\lambda}_i), \nabla_x F(\bm{x}, \vec{\lambda}_0), \bm{x}^*)$
			\EndFor
			\State Save $\{(\bm{x}_i, \vec\lambda_i)\}_{i=1}^N$
			\State Compute $\bm{z}^*_i = \frac{\bm{x}^*_i - \bar{\bm{x}}}{\bm{\sigma_x}}, i=1, ..., N$
			\State $\mathcal{L}(\bm{w}) = \sum_{i=1}^N \text{dist}(g(\bm{w}, \vec{\lambda}_i), \bm{z}_i^*)$
			\Comment{with random restart}
			\State $\bm w^* = OPT(\mathcal{L}, \nabla_w \mathcal{L}, \bm w)$
		\end{algorithmic}
	\end{algorithm}

	\subsection{Direct-training method with back propagation: SOMA BP}
	An adaptive trajectory is a solution that depends on a specific parameter of the physical system, which the optimizer does not control, although it can make use of it. Usually, for many robotics applications, the system does not have an analytical model, thereby preventing direct-learning strategies. However, for quantum dynamics, we show how we can use the model to more directly train the function approximator.  We refer to this approach as SOMA with back propagation (SOMA BP). A sketch of the algorithm is provided in Fig.~\ref{fig:soma_types} (b).
	
	{\begin{figure*}[ht!]
        \centering
        \includegraphics[width=\textwidth]{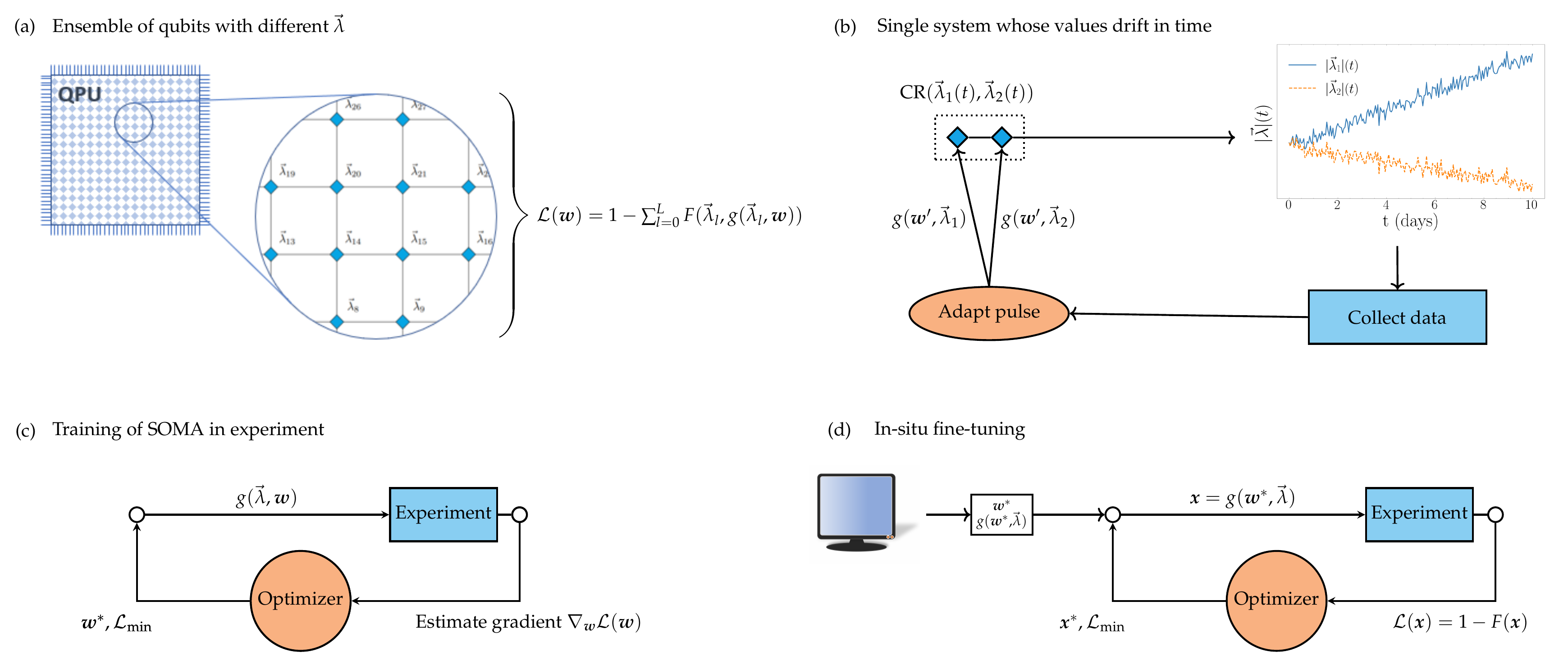}
        \caption{\red{Four diagrams describing two possible use cases of SOMA and two possible experimental implementations of SOMA BP, respectively. (a) We consider a chip with several qubits, each one with its own Hamiltonian parameter $\vec{\lambda}$. We assume parameter variations $\delta \vec{\lambda}$ to be so large that robust pulses are generally ineffective. (b) In the second use case, we consider a single system, the parameters of which vary with time. Pulses are trained on the average cost function over ensembles of qubits. (c) The approximator is trained directly on the experimental setting by using a gradient estimator, in this case a policy gradient \cite{Sutton1999}, which usually requires large numbers of samples to be drawn from the system.} (d) The regressor is first trained first in open-loop simulation and then used as an ansatz for a closed-loop optimization, which leaves the neural-network parameter untouched but modifies the output amplitude parameter (or, alternatively and if possible, the input parameters) to maximize the fidelity for specific experimental configurations. The optimization routine can be freely chosen among gradient-free algorithms, such as Nelder-Mead in Ref.~\cite{Caneva2011}. This method can be a viable option if the experimental setting can be simulated with sufficient precision.}
        \label{fig:experiment_diagram}
    \end{figure*}
    
	Prime examples of adaptive trajectories are the analytical pulses in Sec.~\ref{sec:ansol}. For instance, by using a frequency-dependent solution, DRAG eliminates the leakage inside a qutrit. Moreover, this solution parametrized by the function approximator, $g$, directly depends on the physical system values and can therefore be tuned if these are shifted.  
	
	The cost function for an ensemble of $L$ quantum-optimal-control (QOC) problems defined by problem parameters $\vec \lambda_1, ..., \vec\lambda_L$ is given by:
	\begin{align}\label{eq:unsupervisedloss}
	\mathcal{L}(\bm{w}) = 1 - \frac{1}{L}\sum_{l=1}^L F(g(\boldmath \bm{w}, \vec\lambda_l), \vec\lambda_l) 
	\end{align}
	where $F$, as before, is the figure of merit, e.g., the overlap fidelity of the operation with the target quantum gate \cite{Nielsen2002}.
	
	By parametrizing the solution in terms of a neural network that depends on the gate parameters, gradient-based optimization algorithms can be used to train the network directly off the above cost function. In essence, the usual back propagation of neural networks matches naturally with gradient-descent optimal-control methods such as GRAPE \cite{Khaneja2005}. Thus, while optimizing in the fidelity landscape of the controls, our algorithm is able to simultaneously train the network to adapt to the extraneous system and gate parameters. This method can also be used in combination with a more standard robust-GRAPE approach. In this case, those parameters the calibration and control of which proves difficult, can be excluded from the network input. 

	\begin{algorithm}[H]
		\caption{SOMA BP}
		\label{alg:SOMA2}
		\setstretch{1.20}
		\hspace*{\algorithmicindent} \textbf{Input} $\bm{w},\vec \lambda_0, \vec{\lambda}_{\text{min}}, \vec{\lambda}_{\text{max}}$, optimizer $\text{OPT}$\\
		\hspace*{\algorithmicindent} \textbf{Output} $\bm{w}^*$
		\begin{algorithmic}[1]
			\For{$i=1$ to $L$}
			\State $\vec{\lambda}_i \sim \pi(\vec{\lambda}_{\text{min}}
			, \vec\lambda_{\text{max}})$
			\EndFor
			\State $\mathcal{L}(\bm{w}) = 1 - \frac{1}{L}\sum_{i=1}^L F(g(\bm{w}, \vec{\lambda}_i), \vec{\lambda}_i)$
			\State $\bm{w^*} = OPT(\mathcal{L}, \nabla_w \mathcal{L}, \bm{w})$ \Comment{with random restart}
		\end{algorithmic}
	\end{algorithm}

	\subsection{Experimental adaptation}
	
	\red{Direct experimental application of SOMA is possible both for model-based and model-free implementations. For these purposes, one must have access to a controlled distribution of $\vec \lambda$ values, corresponding to gate parameters, pulse parameters, and system parameters. For \textit{in situ} optimization of the gates, one further requires access to an experimental cost function to ascertain (with low noise and bias) the merit of the pulse sequence. For model-free control learning, the system parameters should still be indexed in some way, e.g., by performing characterization, by using a proxy such as other known characteristics, or by externally tuning parameters (e.g., the qubit frequency via magnetic or Stark shifts). For model-based approaches, one can vary the parameters $\vec \lambda$ in software to map the solution space of the continuous gate sets. Thus, in both cases, provided that there is known variation in some parameters, then one can index them, e.g., discretely in space or (slowly varying) continuously in time, as shown in Fig.~\ref{fig:experiment_diagram} (a) and (b) respectively. 
	
	Whichever parameters for $\vec \lambda$ one chooses for the experiment, the task then becomes to learn the neural-network weights for the maximal performance on the relevant device and gate defined uniquely by $\vec \lambda$. Two different approaches are shown in Fig.~\ref{fig:experiment_diagram} (c) and (d). When an accurate model for the generators of the dynamics is known, then Fig.~\ref{fig:experiment_diagram} (d) is a natural choice whereby offline (i.e.,~open-loop) optimization of the simulated gates is first brought up, and only once the solution class has been learned \textit{in situ} (i.e., closed-loop) is control learning performed. In this secondary step, one can reoptimize either over the space of solutions (fine tuning optimal $\vec \lambda^*$) or over the space of physical controls (fine tuning optimal $\bm{x}^*$).

    Closed-loop optimization directly on the experiment can be performed a number of ways, including numerical and parameter-shift approximations of the gradient, Nelder-Mead \cite{NelderMead1965}, or evolutionary algorithms \cite{Wierstra2014}. In the latter case, for example, Monte-Carlo gradient sampling can be used to estimate the update direction \cite{Shakir2020}:
	\begin{align}
	 \nabla_{\bm{x}}F(\bm{x}) = \frac{1}{\tilde{N}}\sum_{i=1}^{\tilde{N}} F(\bm{x} +  \sigma \bm{\epsilon_i}) \bm{\epsilon_i}
	\end{align}
	where $\epsilon_i \sim \mathcal{N}(\bm{0},\mathbb{I}_Q), \ i=1,..,\tilde{N}$ is a normally distributed stochastic variable sampled $\tilde{N}$ times, $\bm{x}$ is the network output representing the pulse and $\sigma$ is the standard deviation of the sampled pulses. This method is often referred to as natural evolution strategy \cite{Wierstra2014}. Fig.~\ref{fig:experiment_diagram}(c) shows how the experimental gradient of the cost function can be then back propagated via the optimizer (similarly to the GRAPE implementation above) in order to update the network weights $\bm{w^*}$ directly, when the different $\vec \lambda$ can be sampled simultaneously.
	}

	\begin{figure*}[ht!]
		\hspace{-1cm}
		\includegraphics[width=\textwidth]{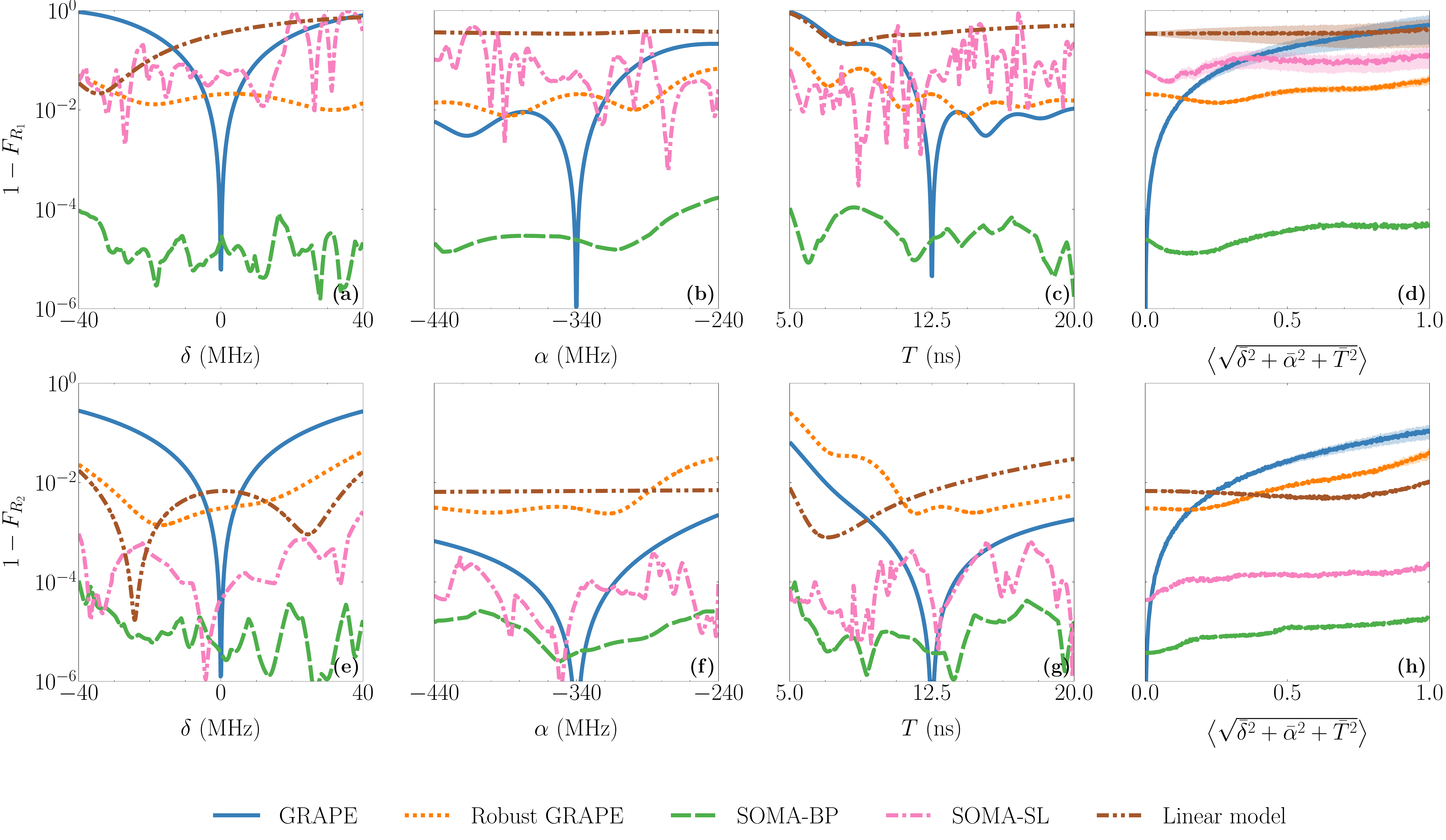}
		\hspace{1cm}
		\caption{The infidelity of pulses predicted by the different optimization methods for the $R_1(\theta=\pi/2)$ gate (first row) and the $R_2(\theta=\pi/2)$ gate (second row) as a function of three different quantum control problem parameters: (a), (e) the detuning $\delta$ between the qubit frequency and the driving frequency; (b), (f) the nonlinearity $\alpha$ of the qutrit; and (c), (g) the total gate duration $T$.  (d) (h) THE average performance (1000 test samples) of the algorithms as a function of the radial distance from the center $\vec{\lambda}_0$ of the parameter space. The terms $\bar{\delta}, \bar{\alpha}, \bar{T}$ indicate that the problem parameters are renormalized to the space $[0,1]^D$ -- see Appendix \ref{appx:methods}.  \red{The shaded regions around each plot line show the standard deviation of the corresponding infidelities.  A more detailed discussion about the standard deviation can be found in Appendix \ref{appx:methods}}. Both gates are optimized with four Fourier components for each one of the two control fields -- see Eq.~\eqref{eq:qutrit_control} --  using  the Hamiltonian in Eq.~\eqref{eq:qutrit_ham_simp}. The range of each parameter is given in Tab.~\ref{tab:table_1qubit_1}. } \label{fig:4plots_1}
	\end{figure*}
	\section{Results} \label{sec:results}
	
	We test our methods and compare to previous optimalßcontrol theoretic approaches. For this purpose, we train our solution networks to learn how to perform continuous gate sets for both single- and two-qubit operations. As a figure of merit of the QOC problems, we choose the gate fidelity \red{ defined in Eq.~\eqref{eq:grapeF}.
    While the gradients of the fidelity with respect to the pulse parameters $x$ can be computed using GRAPE \cite{Khaneja2005, Motzoi2011} (see Section \ref{subsec:grape} and Appendix \ref{appx:backprop}), in the context of these simulations the gradient can also be obtained through automatic differentiation \cite{jax2018github, Rall2006-in}.   }
    To simulate the quantum system, we use a second-order Magnus propagator as derived in Refs.~\cite{Blanes2009, dalgaard2022fast}, and which is compatible with analytical or automatic differentiation \cite{dalgaard2022fast}. For the optimization of all the parameters, we employ the algorithm L-BFGS-B \cite{Liu1989}.
	
	\subsection{Single-qubit gates}
	Single-qubit gates are the fundamental building blocks of quantum circuits. Their most general form -- e.g., as they are implemented in the IBMQ compiler \cite{ibmq} -- is given by: 
	\begin{align}\label{eq:genU}
	&R(\theta_1, \theta_2, \theta_3) =   &\begin{bmatrix}
	\cos(\theta_1/2) &  - e^{i\theta_2}\sin(\theta_1/2) \\
	e^{i\theta_3}\sin(\theta_1/2) & e^{i(\theta_3 + \theta_2)}\cos(\theta_1/2)
	\end{bmatrix}.
	\end{align}
	By choosing the vector parameter $\bm{\theta}$ appropriately, one can construct arbitrary single-qubit unitaries. Therefore, in any optimal quantum control problem, this triple can be considered as a vector of problem parameters, since they define the entire class of QOC problems, the goal of which is the optimization of arbitrary single-qubit gates. 
	
	In the following section we consider an ensemble of QOC problems defined by parameters of the unitary target gate, Hamiltonian parameters, parameters of the control fields, and the evolution time $T$.
	To simplify the problem, we consider a target gate of type
	\begin{align}\label{eq:h_gate}
	R_1(\theta) = \begin{bmatrix}
	\cos(\theta/2) &   \sin(\theta/2) \\
	\sin(\theta/2) & -\cos(\theta/2)
	\end{bmatrix}
	\end{align}
	by setting $\theta_1 = \theta$, $ \theta_2 = \pi$ and $\theta_3 =0$ in Eq.~\eqref{eq:genU}.
	For $\theta = \frac{\pi}{2}$, the gate is the $H$ gate, whereas for $\theta = \pi$ it produces the $X$ gate.\\
	The second family of unitaries that we consider can be obtained by setting $\theta_1 = \pi$ and $\theta_2 = \theta_3 = \theta - \pi$ in Eq.~\eqref{eq:genU}. The resulting family of gates,
	\begin{align}\label{eq:x_gate}
	R_2(\theta) = \begin{bmatrix}
	0 & e^{i (\theta - \pi)} \\
	e^{-i (\theta-\pi)} & 0 \\
	\end{bmatrix}
	\end{align}
	generates, among other unitaries, the $X$ gate for $\theta = \pi$ and the $Y$ gate for $\theta = \frac{\pi}{2}$. \red{We would like to point out that a combination of single-qubit unitary gates as in Eq.~\eqref{eq:x_gate} and Eq.~\eqref{eq:h_gate} is sufficient to generate arbitrary single-qubit unitaries \cite{NielsenChuang2010}.} 
	
	For single-qubit simulations, we consider the Hamiltonian of a superconducting transmon qubit \cite{kjaergaard2020superconducting}. This system can be effectively reduced to a qutrit Hamiltonian \cite{kjaergaard2020superconducting, khani2009optimal}, where the $\ket{0}$ and $\ket{1}$ levels provide the computational subspace and the $\ket{2}$ level represents the leakage.
	\begin{figure*}[ht!]
		\hspace{-1cm}
		\includegraphics[width=\textwidth]{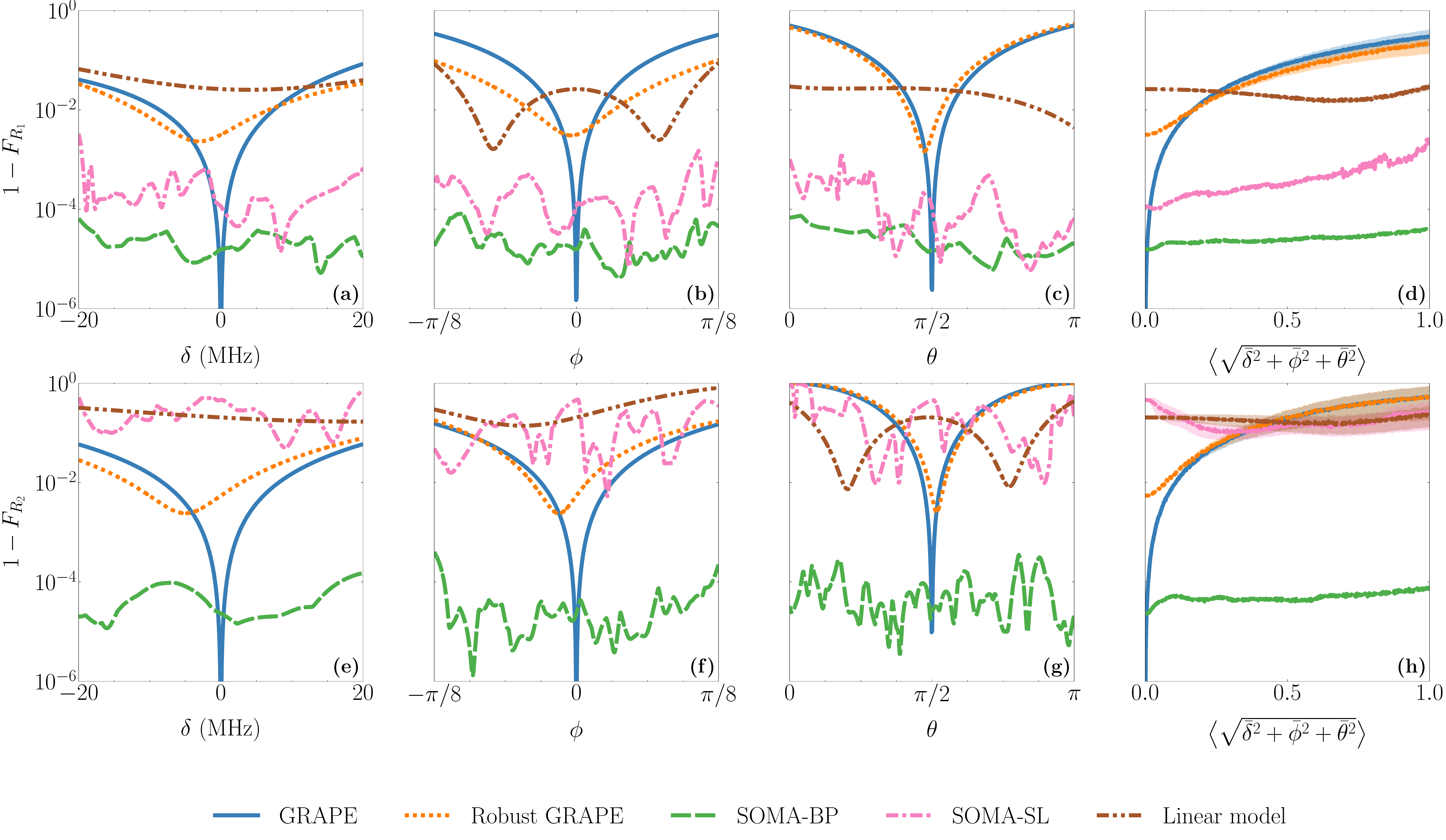}
		\hspace{1cm}
		\caption{The infidelity of pulses predicted by the different optimization methods for the $R_1(\theta)$ gate (first row) and the $R_2(\theta)$ gate (second row) as a function of three different quantum control problem parameters: (a), (e) the detuning $\delta$ between the qubit frequency and the driving frequency; (b), (f) the phase error $\phi$ between the $\sigma_+$ and the $\sigma_{-}$ terms; and (c), (g) the angle $\theta$ parametrizing the target gate. (d), (h) The average performance (1000 test samples) of the algorithms as a function of the radial distance from the center $\vec{\lambda}_0$ of the parameter space. The terms $\bar{\delta}, \bar{\phi}, \bar{\theta}$ indicate that the problem parameters are renormalized to the space $[0,1]^D$ (see Appendix \ref{appx:methods}). \red{The shaded regions around each plot line show the standard deviation of the corresponding infidelities.  A more detailed discussion about the standard deviation can be found in Appendix \ref{appx:methods}.} Both gates are optimized with four Fourier components for each one of the two control fields for $T=10 \ \si{\nano \second}$ using the Hamiltonian in Eq.~\eqref{eq:qutrit_ham_simp}.} \label{fig:4plots_2}
	\end{figure*}
	The drift Hamiltonian for our system \cite{ Motzoi2009} reads
	\begin{align}\label{eq:qutrit_ham}
	&H_d = \omega_d \hat{n} + \alpha \hat{\Pi}_2 \\
	&\hat{\Pi}_j = \ket{j}\bra{j} \\
	&\hat{n} = \sum_j j \hat{\Pi}_j, 
	\end{align}
	where $\omega_d$ is the qubit frequency and $\alpha$ is the anharmonicity.
	Furthermore, we consider a control Hamiltonian of type:

	\begin{align}\label{eq:qutrit_control}
	H_c(t) = \Omega(t) e^{i\phi + i \omega t}\hat{\sigma}_{+} + \Omega(t)^* e^{-i\phi -  i \omega t}\hat{\sigma}_{-},
	\end{align}
	where $\omega$ is the driving frequency and $\phi$ represents a time-independent phase shift between the raising $\hat{\sigma}_{+}$ and the lowering $\hat{\sigma}_{-}$ operators, which can be related to the rotating-wave approximation (RWA) \cite{Cohen-Tannoudji1998, Motzoi2011} and which is, in this model, the only problem parameter influencing the control fields. 
	
	Computing the RWA with detuning $\delta$ allows us to rewrite the drift Hamiltonian as:
	\begin{align}\label{eq:qutrit_ham_simp}
	&H_d = \delta \hat{\Pi}_1 + (\alpha - 2\delta) \hat{\Pi}_2, \\
	&H_c(t) = u_1(t) \hat{X}(\phi) + u_2(t) \hat{Y}(\phi)
	\end{align}
	where $\delta= \omega_d - \omega$, $\hat{X}(\phi) =  e^{i \phi} \hat{\sigma}_{+} +  e^{-i\phi} \hat{\sigma}_{-}$, and $i \hat{Y}(\phi) = e^{i \phi} \hat{\sigma}_{+} - e^{-i\phi} \hat{\sigma}_{-}$.
	
	For the control fields, we employ a Fourier ansatz:
	\begin{align} \label{eq:fourier_ansatz}
	    u_j(t) = \sum_{k=1}^K x_{kj} \sin(\frac{k \pi t}{T}), \ j=1,2
	\end{align}
	with $K$ Fourier modes.
	For the QOC simulations, we set the central values $\delta_0 = 0 \ \si{\giga \hertz}$ and $\alpha_0 = -0.34 \ \si{\giga \hertz}$ \cite{Theis2015}. The parameter vector of the QOC-problem class is given by
	\begin{align}
	\vec{\lambda} = \left(\delta, \alpha, \phi, \theta, T \right)^T. 
	\end{align}	
	For all the single-qubit gate simulations, we consider a multidimensional rectangle centered at $\vec{\lambda}_0 = \left( \delta_0, \alpha_0, \phi_0, \theta_0, T_0 \right)^T$ and with upper bounds defined by $\pm \vec{\lambda}_{\text{max}}$.
	We consider four methods, which allow us to optimize multiple systems simultaneously: standard GRAPE; robust GRAPE, which uses the average GRAPE gradient over an ensemble of QOC-problems in Eq.~\eqref{eq:robust}; a supervised training method, which we refer to as SOMA SL (Algorithm \ref{alg:SOMA1}), using both linear and nonlinear models and where GRAPE solutions are first generated and then approximated via Eq.~\eqref{eq:supervisedloss}; and the unsupervised method, which trains a neural-network pulse directly on the fidelity of an ensemble of QOC problems using back propagation on Eq.~\eqref{eq:unsupervisedloss}. The latter is referred to as SOMA BP (Algorithm \ref{alg:SOMA2}).
	
	In the following section, we consider a QOC problem with $N=500$ time steps of a Magnus-type time integrator, which approximates the unitary temporal evolution of the quantum system \cite{dalgaard2022fast}. Our pulses are parametrized as in Eq.~\eqref{eq:qutrit_control} by $K=4$ Fourier components according to Eq.~\eqref{eq:fourier_ansatz} for each one of the two control fields $\hat{X}$ and $\hat{Y}$. The control fields are multiplied with a scaling factor equal to $1/A_0$ to ensure that the pulse amplitudes do not exceed the typical experimental maximal value of $1\ \si{\giga \hertz}$.
	
	The results are divided into four blocks of four plots each (Fig.~\ref{fig:4plots_1}, Fig.~\ref{fig:4plots_2}), representing models trained on two different sets of problem parameters\red{, the ranges of which are given in Tab.~\ref{tab:table_1qubit_1} and in Tab.~\ref{tab:table_1qubit_2}.} Here we limit the number of different parameters to three, although larger numbers are possible depending on the parameter range taken into account, the specifics of the physical system, the size of the neural network, and the number of systems sampled (see Appendix \ref{appx:methods}). The first three plots in each row show the infidelity as a function of one varying parameter, while the other parameters are kept at their original values given by $\vec{\lambda}_0$. The fourth plot shows the average performance of the different methods as a function of the radial distance from the initial QOC problem $\vec{\lambda}_0$. This last plot ensures that the performance of the methods is stable when different parameters are changed simultaneously across the entire range of the parameter space. 
	
	In Fig.~\ref{fig:4plots_1}, the problem parameters $\delta$ (the qubit-drive detuning), $\alpha$ (the nonlinearity of the qutrit system), and $T$ (the gate evolution time) and the target gates $R_1(\theta)$ defined by Eq.~\eqref{eq:h_gate} -- first row -- and $R_2(\theta)$ defined by Eq.~\eqref{eq:x_gate} -- second row -- are considered. The blue continuous line shows the GRAPE solution for the $\vec{\lambda}_0$ parameter, whose fidelity shows an exponential decay as a function of the distance from the center of the parameter space. The orange dashed line shows the performance of robust GRAPE and the brown dashed line the performance of SOMA SL with a linear model, both with infidelities around $10^{-2}$. The pink dashed line shows the performance of SOMA SL with a neural network and the green dashed line the performance of SOMA BP.
	We observe that for the second gate  $R_2(\theta)$, both methods are able to deliver an average infidelity below $10^{-4}$, whereas for the first one $R_1(\theta)$, SOMA BP clearly outperforms SOMA SL.
	
	In Fig.~\ref{fig:4plots_2}, the problem parameters are as follows: drive frequency detuning $\delta$, the phase mismatch in the control fields $\phi$, and the gate angle $\theta$, together with the target gates  $R_1(\theta)$ defined by Eq.~\eqref{eq:h_gate} -- first row -- and $R_2(\theta)$ definied by Eq.~\eqref{eq:x_gate} -- second row. We note the same general trends as in Fig.~\ref{fig:4plots_1}. In particular, we observe that although both SOMA methods are closer in terms of performance for  $R_1(\theta)$ with a little advantage for SOMA BP, for gate $R_2(\theta)$ SOMA BP clearly outperforms SOMA SL.
	We can again note that SOMA BP mostly outperforms all the other methods, providing pulses that are stable over a large range of parameters. 
	For simulations with single-qubit gates, we use a two-layered network with 256 components per layer and eight neurons in output -- the output space of the approximator has an output dimension $Q = 2K$.
	\begin{table}
		
		\resizebox{0.9\columnwidth}{!}{%
			
			\begin{tabular}{c*{6}{>{$}c<{$}}}
				& \text{$\delta (\si{\mega \hertz})$}        & \text{$\alpha (\si{\mega \hertz})$}       & \text{$\phi$}       & \text{$\theta$} & \text{$1/A_0$}   & \text{$T (\si{\nano \second})$}     \\
				\hline
				Center  & 0 & -340 & 0 & \frac{\pi}{2} & 0.01 & 10 \\
				Maximum  & 40 & -240 & 0 & \frac{\pi}{2}  & 0.01 & 20 \\
				Minimum  & -40 & -440 & 0 & \frac{\pi}{2}  & 0.01 & 5 \\
			\end{tabular}
		} 
		\caption{The parameter range for the optimization of the single-qubit gates $R_1(\theta)$ and $R_2(\theta)$ as given in Fig.~\ref{fig:4plots_1}. For this simulation, we do not vary the angle parameters of the gate, but, rather, the physical parameters of the qutrit, such as $\delta$ and $\alpha$ and the evolution time $T$.}
		\label{tab:table_1qubit_1}
	\end{table}	
	\begin{table}
		
		\resizebox{0.9\columnwidth}{!}{%
			
			\begin{tabular}{c*{6}{>{$}c<{$}}}
				& \text{$\delta (\si{\mega \hertz})$}        & \text{$\alpha (\si{\mega \hertz})$}       & \text{$\phi$}       & \text{$\theta$}  & \text{$1/A_0$}   & \text{$T (\si{\nano \second})$}     \\
				\hline
				Center  & 0 & -340 & 0 & \frac{\pi}{2} & 0.01 & 10 \\
				Maximum  & 20 & -340 & \frac{\pi}{8} & \pi & 0.01 & 10 \\
				Minimum  & -20 & -340 & -\frac{\pi}{8} & 0 & 0.01 & 10 \\
			\end{tabular}
		} 
		\caption{The parameter range for the optimization of the single-qubit gates $R_1(\theta)$ and $R_2(\theta)$ as given in Fig.~\ref{fig:4plots_2}. In this case, we not only vary the detuning but also the angles of the gates and the phase factor $\phi$.}
		\label{tab:table_1qubit_2}
	\end{table}	
	\subsection{CR gate with leakage}
	\begin{figure*}[ht!] 
		\hspace{-1cm}
		\includegraphics[width=\textwidth]{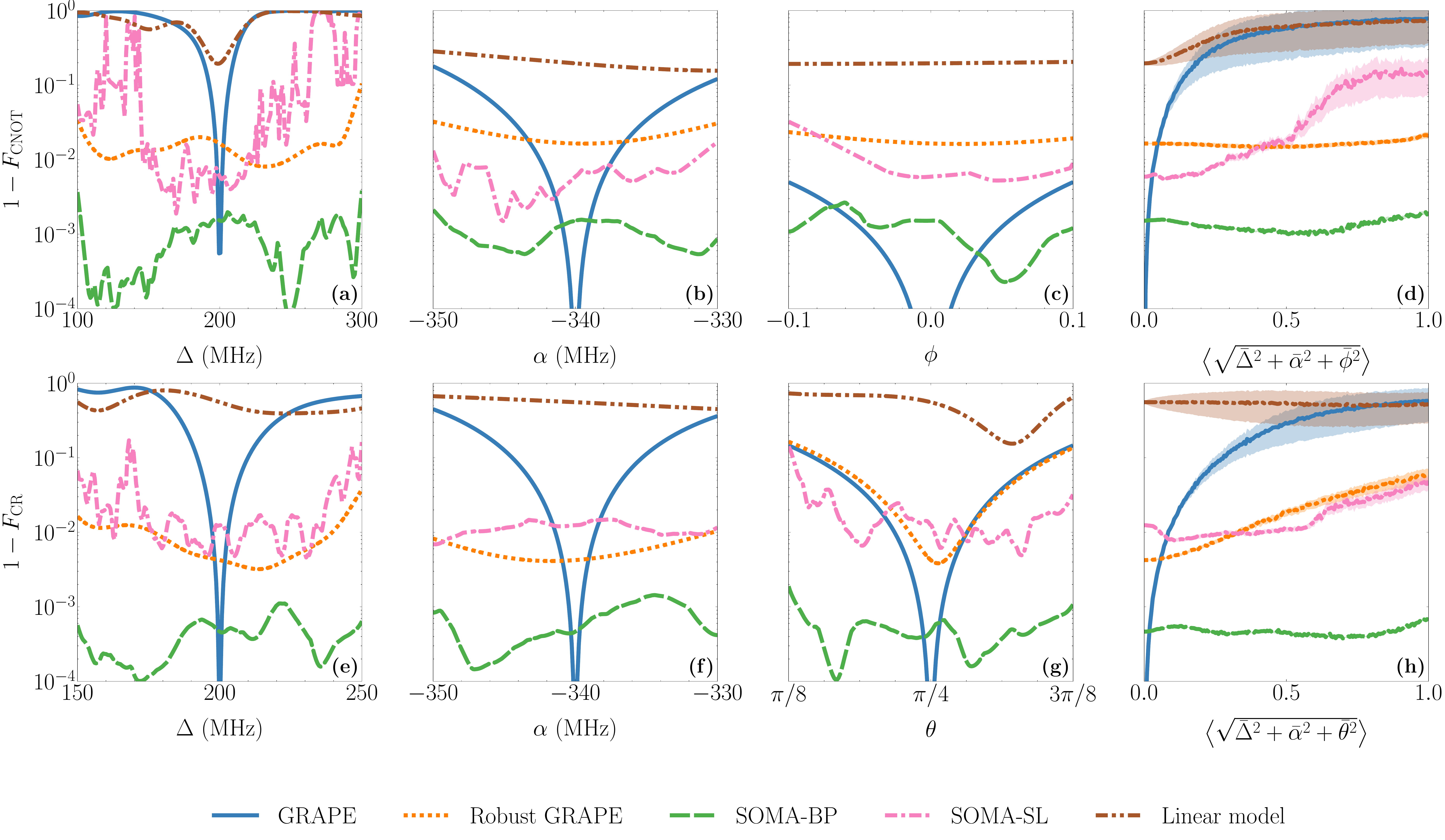}
		\hspace{1cm}
		\label{fig:CR_plots_1}
		\caption{The infidelity of pulses predicted by the different optimization methods for the direct controlled-NOT (CNOT) gate (first row) and the $CR(\theta)$ gate (second row) as a function of different quantum control problem parameters: (a), (e) the frequency difference $\Delta$ between the two qubits; (b), (f) the nonlinearity $\alpha$ of the qutrits; (c) the phase error $\phi$ between the control fields $\hat{\sigma}_{+}$ and $\hat{\sigma}_{-}$; (g) the angle $\theta$ parametrizing the CR continuous family of gates given by Eq.~\eqref{eq:cr_theta}. Plots (d) and (h) show the average performance (1000 test samples) of the algorithms as a function of the radial distance from the center $\vec{\lambda}_0$ of the parameter space. The terms $\bar{\Delta}, \bar{\alpha}, \bar{\phi}, \bar{\theta}$ indicate that the problem parameters are renormalized to the space $[0,1]^D$ -- see Appendix \ref{appx:methods}. \red{The shaded regions around each plot line show the standard deviation of the corresponding infidelities.  A more detailed discussion about the standard deviation can be found in Appendix \ref{appx:methods}}. Both gates are optimized with 30 Fourier components for each one of the four control fields and for total gate duration $T=90\ \si{\nano \second}$.} \label{fig:2qubit_4plots_2}
	\end{figure*}
	The cross-resonance (CR) gate is a two-qubit gate activated by microwave fields, which drive one of the qubits (target) at the frequency of the other (control). The gate is implemented in the context of quantum computing with superconducting qubits \cite{Paraoanu2006, Rigetti2010, Kirchhoff2018, Malekakhlagh2022Jan}. The gate gives rise to a $ZX$-type interaction \cite{Paraoanu2006}, which can then be used to generate different two-qubit entangling unitaries. 
	The CR gate can be embedded in a higher-dimensional system, e.g., where two qutrits are capacitively coupled together. In this case, the target gate is a two-qubit gate, but the unitary generated by the Hamiltonian evolution is a two-qutrit gate, thus the fidelity is only computed with respect to the computational subspace -- the CR gate is constructed using the $\ket{0}$ and $\ket{1}$ levels. The Hamiltonian of a two-transmon system reads \cite{Rigetti2010}
	\begin{align}
	&H(t) = H_d + H_c(t) \nonumber\\
	&H_d = \sum_{j=1}^2 \left( \omega_j \hat{b}_j^{\dagger}\hat{b}_j + \alpha_j \hat{b}_j \hat{b}_j^{\dagger} \hat{b}_j^{\dagger}\hat{b}_j \right) + J\left(\hat{b}_1\hat{b}_2^{\dagger} + \hat{b}_1^{\dagger}\hat{b}_2\right)\nonumber\\
	&H_c(t) = \sum_{j=1}^2 \Omega(t) e^{i \omega_j t + i\phi} \hat{b}_j^{\dagger} + \Omega(t)^* e^{-i \omega_j t - i\phi} \hat{b}_j,
	\end{align}
	where $J$ is the coupling strength of the transmon-transmon interaction, $b_j$ is the lowering operator on the $j$th transmon, $\Omega$ is the driving field, and a Duffing-oscillator approximation is performed \cite{khani2009optimal}.
	A further standard RWA allows us to simplify the problem, introducing at the same time the notation from Eq.~\eqref{eq:qutrit_ham}, such that
	\begin{align}
	H_d =& \ \Delta \hat{n}_1 + \alpha (\hat{\Pi}^{(1)}_2 + \hat{\Pi}^{(2)}_2) + J\left(\hat{b}_1\hat{b}_2^{\dagger} + \hat{b}_1^{\dagger}\hat{b}_2\right) \\
	H_c(t) =& \sum_{j=1}^2 \bigg( u^{(j)}_1(t) \left(e^{i\phi} \hat{b}_j^{\dagger} +  e^{-i\phi} \hat{b}_j \right)+ \\ \notag &u^{(j)}_2(t) \left(e^{i\phi} \hat{b}_j^{\dagger} -  e^{-i\phi} \hat{b}_j\right) \bigg),
	\end{align}
	where $\Delta = \omega_1 - \omega_2$.
	As control operators, we use the projectors $\hat{X}_j(\phi) =  e^{-i\phi}\hat{b}_j +  e^{i\phi}\hat{b}_j^{\dagger}$ and $i \hat{Y}_j (\phi) = e^{i\phi} \hat{b}_j^{\dagger} -  e^{-i\phi} \hat{b}_j$   both on the control qutrit ($j=1$) and the target qutrit ($j=2$), as in Eq.~\eqref{eq:qutrit_control}, such that
	\begin{align}
	    &H_c(t) = \sum_{j=1}^2 u^{(j)}_1(t) \hat{X}_j(\phi) + u^{(j)}_2(t) \hat{Y}_j(\phi)
	\end{align}
	where $u^{(1)}_j(t),\ u^{(1)}_j(t)$ are the control fields on the $j$th qutrit (for the $\hat{X}_j$ and the $\hat{Y}_j$ operators respectively). Note that all the fields operate at frequencies near-resonant to the target qubit.
	The Hamiltonian parameters are centered at values $\Delta_0 = 0.2 \ \si{\giga \hertz}$, $\alpha_0 = - 0.34 \ \si{\giga \hertz}$, $J_0 = 0.01 \ \si{\giga \hertz}$.
	For each control field, we use a Fourier parametrization \cite{Caneva2011, Kirchhoff2018}:
	\begin{align}\label{eq:2qutrit_control}
	    \forall j=1,2, \ \forall i=1,2: \ u_i^{(j)}(t) = \sum_{k=1}^K x^{i,j}_{k} \sin (\frac{k \pi t}{T})
	\end{align}
	
	We further assume, as in the single qubit case -- see Eq.~\eqref{eq:qutrit_control}, that the control fields are influenced by a phase factor $\phi$, which then counts as a QOC-problem parameter in the two-qubit simulations.
	
	\begin{table}
		
		\resizebox{\columnwidth}{!}{%
			
			\begin{tabular}{c*{6}{>{$}c<{$}}}
				& \text{$\Delta (\si{\mega \hertz})$}        & \text{$\alpha (\si{\mega \hertz})$}       & \text{$J (\si{\mega \hertz})$}   & \text{$\phi$}  & \text{$A_0$}     & \text{$T (\si{\nano \second})$}     \\
				\hline
				Center  & 200 & -340 & 10 & 0 & 0.05 & 90 \\
				Maximum  & 300 & -330 & 10 & 0.1 & 0.05 & 90 \\
				Minimum  & 100 & -350 & 10 & -0.1 & 0.05 & 90 \\
			\end{tabular}
		} 
		\caption{The parameter range for the optimization of the two-qubit CNOT gate. The range of values $\Delta$, $\alpha$ and $\phi$, as well as the evolution time $T$, are determined by considering typical experimental settings of state-of-the-art superconducting quantum circuits \cite{Malekakhlagh2022Jan, Malekakhlagh2022Oct, Heya2021}. For the value of $\phi$, we assume a small phase error influencing the $\hat{X}$ and $\hat{Y}$ control fields both on the target and the control qubits.}
		\label{tab:CNOTpar}
	\end{table}

	\begin{table}
		
		\resizebox{\columnwidth}{!}{%
			
			\begin{tabular}{c*{7}{>{$}c<{$}}}
				& \text{$\Delta (\si{\mega \hertz})$}        & \text{$\alpha (\si{\mega \hertz})$}       & \text{$J (\si{\mega \hertz})$}       & \text{$\phi$}  & \text{$\theta$} & \text{$A_0$}     & \text{$T (\si{\nano \second})$}     \\
				\hline
				Center  & 200 & -340 & 10 & 0 & \frac{\pi}{4} & 0.05 & 90 \\
				Maximum  & 250 & -330 & 10 & 0 & \frac{\pi}{8} & 0.05 & 90 \\
				Minimum  & 150 & -350 & 10 & 0 & \frac{3\pi}{8} & 0.05 & 90 \\
			\end{tabular}
		} 
		\caption{The parameter space for the optimization of the two-qubit $\text{CR}$ gate as given in Eq.~\ref{eq:cr_theta}. Compared to the simulation of the $\text{CNOT}$ gate, given in Tab.~\ref{tab:CNOTpar}, the range of the target-control frequency detuning $\Delta$ is reduced but the continuous parameter $\theta$ is also considered.}
		\label{tab:CRpar}
	\end{table}
		\begin{figure*}[ht!] 
		\centering
		\hspace{-0.5 cm}
		\resizebox{18cm}{!}{\includegraphics[width=\textwidth]{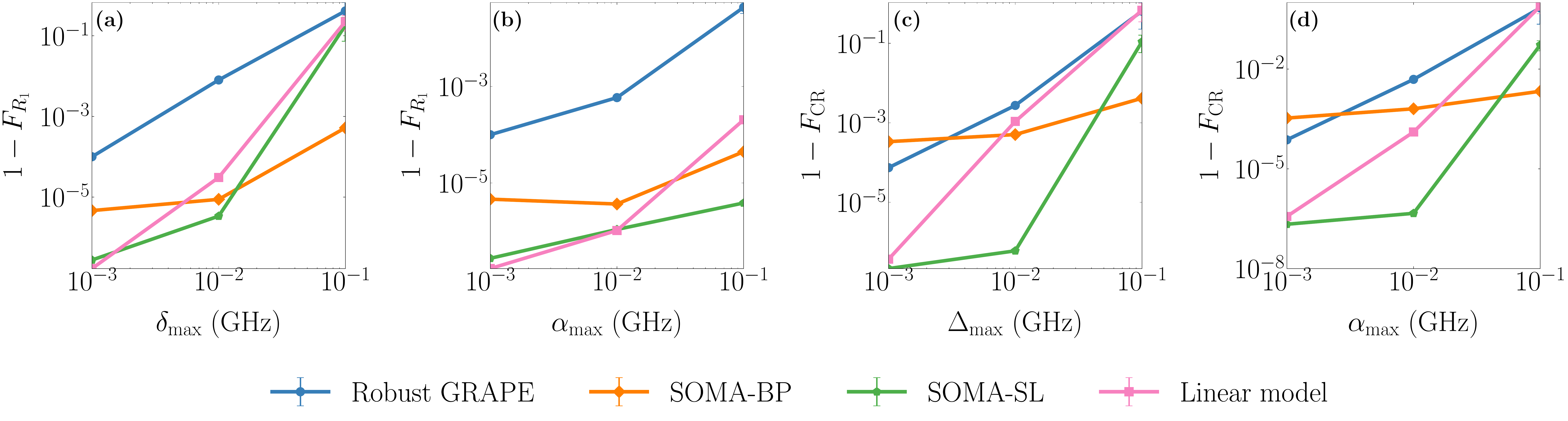}}
		\hspace{0.5 cm}
		\caption{A comparison of four different generalized optimization methods as a function of the parameter space size, when only one parameter range is varied, while all the others are kept constant at $10^{-3} \ \si{\giga \hertz}$: (a), (b) the results for the single-qubit gate $R_1(\theta)$; (c) and (d) the results for the $\text{CR}(\theta)$ gate. Both axes are on a logarithmic scale and represent the average infidelity over 1000 test samples. The values $\delta_{\text{max}}, \alpha_{\text{max}}$, and $ \Delta_{\text{max}}$ on the horizontal axes represent the maximum range of the corresponding problem parameter $\delta, \alpha, \Delta$. The maximum range defines the parameter space over which the QOC problems are sampled. We observe how the neural network trained with the supervised algorithm outputs pulses with higher average fidelity than the other methods but it fails nonetheless when the detuning variation in the two different systems is increased up to the order of 100 \si{\mega \hertz}. In this case, the model trained with SOMA BP, however, still outputs high-fidelity pulses.} \label{fig:rob_plot}
	\end{figure*}
	For two-qubit gates with leakage we present two different simulations, one for the CNOT gate alone and one for a family of CR-like gates with the following parametrization:
	\begin{align}\label{eq:cr_theta}
	\text{CR}(\theta) = \exp[i\theta (Z \otimes X)]
	\end{align}.
	
	For the two-qutrit gates we use a two-layered neural network with 500 neurons per layer and 120 neurons in output. Here, the output of the approximator has a dimensionality $Q = 4K$, where $K=30$ is the number of Fourier frequencies for each one of the four control fields -- see Eq. \eqref{eq:2qutrit_control}. The time evolution is given by 5000 Magnus steps.

	The results of the pulse class learning are shown in Fig.~\ref{fig:2qubit_4plots_2}.
    The evolution time is chosen as $T= 90 \  \si{\nano \second}$, an improvement of about a factor of 2 over typical experimental durations. The first row of plots shows the results for the CNOT gate, where variations of the parameters $\Delta$ (the frequency difference between the control and target qubits), $\alpha$ (the nonlinearity of the qutrits), and $\phi$ (the phase term of the control fields) are considered. The second line shows the results for the $\text{CR}(\theta)$ gate as defined in Eq.~\eqref{eq:cr_theta}. The maximal range for each parameter for each one of the two gates is described in Tab.~\ref{tab:CNOTpar} and Tab.~\ref{tab:CRpar} respectively.  We observe that, for both gates, the neural-network pulse trained with SOMA BP outperforms the other algorithms. For the CNOT gate, it can produce pulses that are robust over a very large detuning range ($\vert \Delta_{\text{max}} - \Delta_{\text{min}} \vert = 200 \ \si{\mega \hertz}$) by training on just 100 different systems sampled from a uniform distribution. For the CR gate, we choose a range of $100\ \si{\mega \hertz}$, while also taking into account the dependence on the angle $\theta$. It is possible that a very large amount of samples or long training times could improve the performance of SOMA SL further; however as the size of the Hilbert space increases, the optimization of large sample numbers on classical machines can become computationally too time consuming.

	\section{Performance analysis} \label{sec:robvsada}
	In this section, we compare the different meta-optimization methods as the size of the QOC problem-parameter space increases. In particular, we vary the order of magnitude of the laser-frequency detuning $\delta$, the qubit-frequency detuning $\Delta$ for the two-qubit gate, and the nonlinearity $\alpha$ for both the single-qubit gate and the CR gate. For the sake of this analysis, we only vary the maximal range of one single parameter at a time, while the other problem parameters have a fixed maximal range of $10^{-3} \ \si{\giga \hertz}$. As hyperparameters for the different algorithms, we use the same values that have proved to be effective in the previous simulations. For SOMA SL, we use up to 10000 sample problems optimized with GRAPE, whereas for SOMA BP we optimize the average fidelity with 500 system samples for the single-qubit gate and 100 systems for the two-qubit gate.
	
	The results are shown in Fig.~\ref{fig:rob_plot}. We observe that although supervised training using the minima produced by GRAPE shows lower infidelities for small variations of the detuning (up to 10\ \si{\mega \hertz}), it fails when this value is increased to 100 \si{\mega \hertz}, whereas the neural network trained with back propagation of the fidelity still produces valid optima of the QOC problem. In particular, we observe a crossing between 10 and 100 MHz for both types of detuning ($\delta$ and $\Delta$) and for the nonlinearity $\alpha$ in the two-qutrit case, where the performance of the supervised method (SOMA SL, green line) worsens dramatically, whereas SOMA BP (orange line) is able to keep the fidelity at high, experimentally valid values. \red{In general, we observed that SOMA SL significantly outperforms SOMA BP for small parameter variations, where the precision of the non-linear regression is high enough to reproduce the pulse variations perfectly (see also Sec.~\ref{fig:rob_plot}). As a consequence, SOMA SL can be a useful tool to achieve adaptive robustness against small parameter variations, where it clearly surpasses SOMA BP, whereas the latter shines when the parameter variations and the number of Fourier components are comparatively larger. Furthermore, SOMA BP is capable of handling large variations of multiple parameters at the same time, as shown in Fig.~\ref{fig:4plots_2}. In all cases, at least one neural-network approach outperforms robust GRAPE and the linear regression model significantly.
	We argue that sampling large amounts of problem parameters could actually lead to a better performance of SOMA SL for \textit{in-situ} physical systems or numerical simulations where sampling proves fast and efficient. However, as we show in Fig.~\ref{fig:sample_size1q}, it seems that for SOMA SL (green line) there exist systems where its improvement as a function of the sample size is limited, which makes SOMA BP (orange line) a better option, if its implementation is possible. Likewise robust GRAPE (blue line) and the linear model (pink line) gain limited benefit from increasing sample sizes. In particular for SOMA SL, we can often observe behaviors such as branching and outliers in the training data set, which probably contribute to a loss in the quality of the approximations. One may try to increase the precision of the model by using loss functions that are sensitive to outliers, such as the Huber loss \cite{Huber1964}, or to restrict the use of SOMA SL to systems where limited parameter drifting does not prevent the regressor from learning a high-quality representation of the solution space.  

	We also study whether varying the chosen samples during training (i.e., minibatching) can affect the performance of our algorithms; in particular for robust GRAPE and SOMA BP. Computing the gradient over a batch of samples gives rise to a batched version of the algorithms. More specifically, a batched version of robust GRAPE, called bGRAPE, has been studied in Ref.~\cite{Wu2019}, where impressive robustness is achieved by combining batches of problem parameters with momentum-based stochastic gradient descent. This is, of course, a different algorithm than L-BFGS-B, i.e., the optimization algorithm employed in all simulations discussed in this paper, and it does not similarly guarantee near-quadratic convergence \cite{Liu1989}. For SOMA SL, computation of the gradient based on the MSE loss over a batch of samples leads to standard neural-network training with stochastic gradient descent. For the systems we consider, we do not observe an improvement of bGRAPE over robust GRAPE (called sGRAPE in Ref.~\cite{Wu2019})  neither with L-BFGS-B nor with ADAM \cite{Kingma2015AdamAM}. We believe that this is due to the use of Fourier components, which allow for more controllability of the quantum system \cite{Motzoi2011}, and the use of curvature information granted by both algorithms. Nonetheless, exploration of the effect of varying samples during training remains an interesting perspective worthy of further studies and commitment, both in the context of adaptive and robust control.}
	
		\begin{figure}
		\resizebox{6.5cm}{6.0cm}{\includegraphics[width = 6 cm]{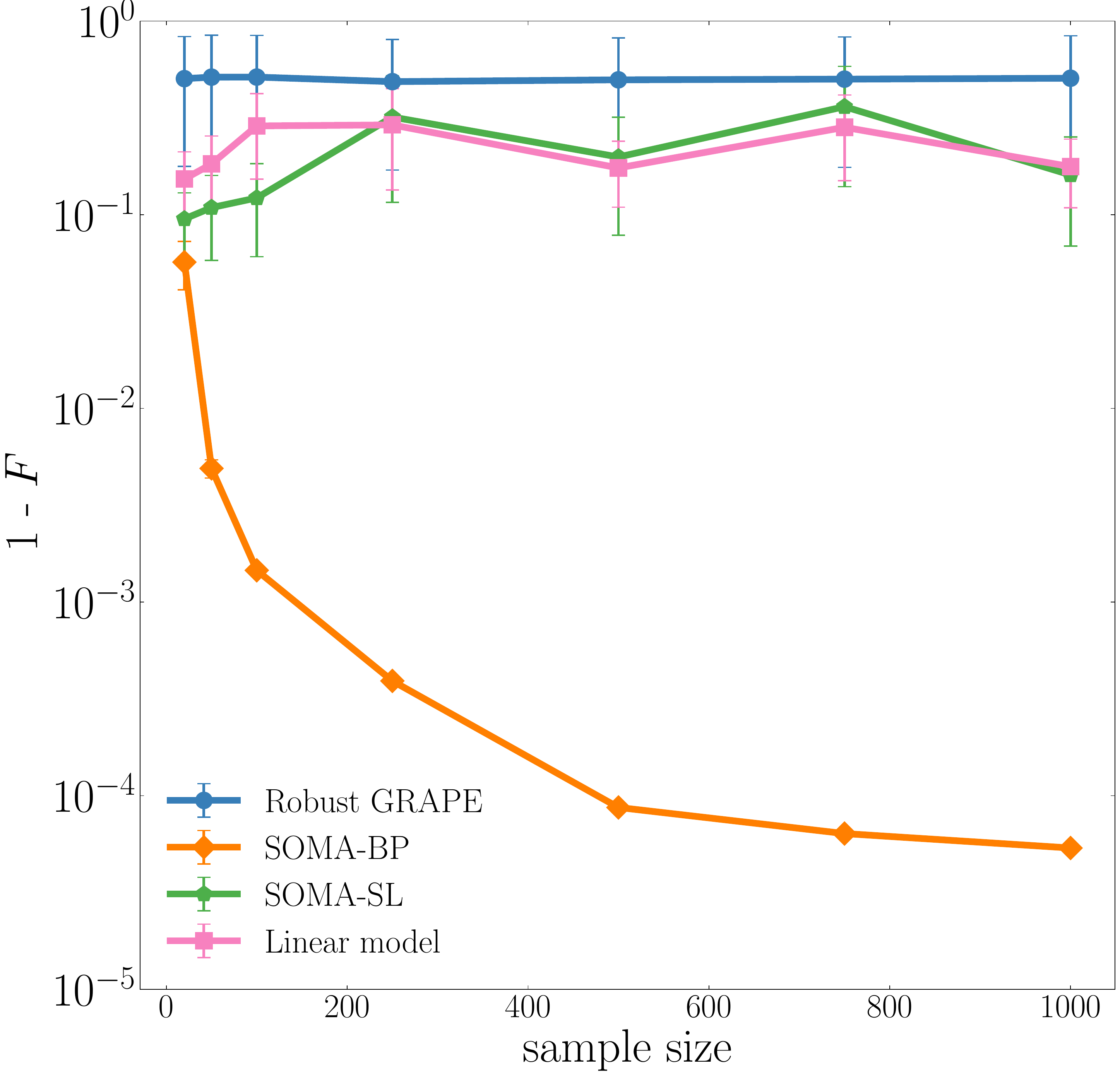}}
		\caption{An example of the average performance (1000 test samples) of robust GRAPE, SOMA SL, SOMA BP and the linear model as a function of the number of (training) systems sampled for the single-qubit $R_2(\theta)$ gate with the same parameters as in Fig.~\ref{fig:4plots_2}. We observe here how in this case the average infidelity in SOMA BP decreases as a function of the sample size, whereas the other methods show little to no improvement.}
		\label{fig:sample_size1q}
	\end{figure}
	
	\red{In the last part of the analysis, we discuss the scaling of the network approximations as we increase the output dimension, the number of qubits and the input dimension. Since we aim at controlling single- and two-qubit gates, the number of expected controls only scales linearly with the number of qubits; and since the weak coupling between qubits drops off roughly exponentially with distance, we also do not expect the search complexity in state space to increase dramatically. In simulations, we expect both algorithms to behave similarly to GRAPE (or a different gradient-based control algorithm, if this is implemented) as the dimension of the Hilbert space increases. This is due to the underlying time evolution, which in both cases is given by the Trotterization or, as in our case, the Magnus expansion. The approximation of the quantum propagator affects the gradient-based optimization for single QOC problems, which generates the target data for SOMA SL, but also acts as an activation function for the neural network \cite{Wu2019} in SOMA BP.
	As a consequence, we do expect the state space and the equations of motion to scale exponentially with the number of qubits, which is a general problem for all control and compilation tasks. Just as in the general case, we expect a combination of informed state-space parametrizations (such as tensor networks and sparse algebras) and of quantum-aided optimization (as in Sec.~\ref{sec:cgl}) largely to address this important problem. Moreover, by considering increasing numbers of qubits, the number of pulse parameters and control fields increases consequently. However control fields acting on different qubits usually commute, which can dramatically decrease correlations between the pulse components. Therefore, one does not need, in general, to have a single neural network outputting all pulse parameters at once, but, rather, several different neural networks, one for each group of control fields, which commute between each other. For SOMA SL, we need to generate a large data set of QOC-problem solutions by means of a standard quantum control algorithm, which may be slow for many-qubit systems. However, this task can be easily parallelized, since the different optimizations are independent of each other. In this case, the main obstacle is not represented by the nonlinear regression over the data, but, rather, by the quality of the data generated by the quantum control algorithm for each solution in the parameter space. Since the problem is high dimensional, the solution space will probably exhibit structures such as branching and outliers that are difficult to include in the nonlinear regression. A possible option here is to employ algorithms for data reduction and clustering, in order to obtain a high-quality representation of the solution space. 
    As for SOMA BP, one may distinguish between simulation and experimental implementation. In the former case, the evaluation of the infidelity and its gradient as a function of the time evolution represents the main bottleneck -- see Tab.~\ref{tab:alg_features} in Appendix \ref{appx:methods} -- together with vanishing gradients, which are also a well-known problem for other NISQ use cases and represent one of the main obstacles to any experimental application. Vanishing gradients could also be tackled with alternatives to back propagation (see, e.g., Ref.~ \cite{Lee2014diff}). Finally, for large output spaces and very deep networks, GPU training and stochastic gradient-descent algorithms may provide useful speedup, as it is the standard in deep learning. \\
	} 
	
	\section{Conclusion}\label{sec:conc}
	In this work, we show how to engineer solutions of problems in quantum optimal control that depend on problem parameters located outside the optimization routine. This includes physical system parameters, other external parameters such as the pulse time or bandwidth, and gate parameters such as rotation angles.
	
	We therefore propose two methods to learn large classes of quantum gate-synthesis problems, SOMA SL and SOMA BP. We show through experimentally relevant examples that these methods prove able to learn adaptive solutions to generalized QOC problems. The output gates have fidelities that remain very high over the entire continuous parametrization of the gate sets, for typically large ranges as would be encountered experimentally.  These continuous gate sets provide the opportunity to be used as computational primitives in compilation tasks, in NISQ variational algorithms, and for entire arrays of qubits rather than individually optimized ones.  
	
	\section*{Acknowledgements}
	We thank Michael Schilling, Boxi Li, Martino Calzavara and Markus Schmitt for  useful discussions.
	We acknowledge support from the German Ministry for Education and Research, under the QSolid project, Grant No.~13N16149,  the Deutsche Forschungsgemeinschaft
    (DFG, German Research Foundation) under Germany’s
    Excellence Strategy -- Cluster of Excellence Matter and
    Light for Quantum Computing (ML4Q) EXC 2004/1 --
    390534769, the Jülich Supercomputing Center, and from AIDAS: AI, Data Analytics and Scalable Simulations, which is a Joint Virtual Laboratory gathering the Forschungszentrum Jülich (FZJ) and the French Alternative Energies and Atomic Energy Commission (CEA). The simulations were realized using the JAX \cite{jax2018github} and FLAX \cite{flax2020github} libraries.

	\bibliography{bibliography}
	
	\appendix
	
	\section{Gradients of the fidelity}\label{appx:backprop}

	GRAPE \cite{Khaneja2005, Defouquieres2011} is the standard open-loop gradient-based method for QOC solutions. It can be implemented together with a Magnus-based propagator -- see \cite{dalgaard2022fast}.
	The gradient can then be used in combination with a gradient-based optimization algorithm, e.g., the L-BFGS-B algorithm \cite{Byrd1994, Liu1989}. Variations of GRAPE exist which exploit the advantages of parametrizing the pulse according to a specific set of basis functions (e.g. Fourier basis) \cite{Motzoi2011} or compute the higher order derivatives of the fidelity with respect to the pulse \cite{Defouquieres2011, dalgaard2020hessian}. In this paper, GRAPE is always used in combination with L-BFGS-B, which uses fast Hessian estimation \cite{2020SciPy-NMeth}.
	
	For a given set of parametrized functions $s_k: t \mapsto s_k(t),\ k=1,...,K$ describing the time dynamics of the control fields $u_j(t)$ with control parameters $x_{kj}$ -- see Eq.~\eqref{eq:qutrit_control} and Eq.~\eqref{eq:2qutrit_control} --, and which can be time-sliced in values $s_{ki}=s_k(t_i), i=0,...,N_{\text{evo}}$, the gradient of the cost function with respect to the control parameters can be computed using the chain rule \cite{Motzoi2011}:
	\begin{align}\label{eq:group}
	\pdv{F}{x_{kj}} =  \pdv{u_{ij}}{x_{kj}} \pdv{F}{u_{ij}} 
	\end{align}
		where the $k$ index runs over the number of basis functions components, the $i$ index over the time-slice and the $j$ index over the different control operators. In a similar way, this can be applied to a neural-network parametrization $g: \mathbb{R}^D \mapsto \mathbb{R}^Q, \vec{\lambda} \longrightarrow g(\vec \lambda)$ of the GRAPE pulse, mapping a given number of meta-parameters to the pulse space, the original formula can be rewritten to output the gradients of the fidelity with respect to the neural-network parameters $\left( w_{ml},\ b_{l} \right)$:
	\begin{align}
		&\pdv{F}{w_{ml}} = \pdv{x_{kj}}{w_{ml}} \pdv{u_{ij}}{x_{kj}} \pdv{F}{u_{ij}} \\
		&\pdv{F}{b_{l}} = \pdv{x_{kj}}{b_{l}} \pdv{u_{ij}}{x_{kj}} \pdv{F}{u_{ij}} 
	\end{align}
	with the same indexing as Eq.~\eqref{eq:group}. In this case, the neural-network output values, which give the coefficients of the time-dependent basis functions, i.e., the terms $x_{ik} = g(\vec{\lambda})_{ik}$ depend on the QOC problem parameters $\vec{\lambda}$. 
	
	\section{Implementation details}
	\label{appx:methods}

	\begin{table*}[ht!]
	\begin{tabular}{|c||c|c|}
         \hline
         \multicolumn{3}{|c|}{Comparison of SOMA methods} \\
         \hline
         Algorithm parameter & SOMA SL & SOMA BP  \\
         \hline
         QOC cost function evaluations   & $n_{\text{fev}} \cdot N$   &  $n_{\text{fev}} \cdot L \cdot N_{\text{guesses}} $ \\
         Computation time &   $t_{\text{GRAPE}} \cdot N + t_{\text{REG}}$ & $t_{\text{BPROP}} \cdot L$ \\
         Number of samples  & $N$ & $L$\\
         \hline
    \end{tabular}
    \caption{Comparison of SOMA SL and SOMA BP in terms of their performance features. $n_{fev}$ and $t_{\text{GRAPE}}$ are here the maximal number of function evaluations and computation time required by L-BFGS-B with GRAPE to solve a single quantum control problem. $t_{\text{REG}}$ is the time required for the neural-network regression. $t_{\text{BPROP}}$ is the time required by the neural network back propagation for one sample.}
    \label{tab:alg_features}
    \end{table*}
    
	For every physical system considered -- single qubit gate with leakage and two-qubit gate with leakage, each one with different target gates and system parameters -- we first define an interval for each parameter. Values are sampled from a uniform distribution over the hyper-volume defined by the intervals of parameters. The system together with the interval of parameters defines a family of QOC problems, which we analyze with one of 3 methods: Robust control with GRAPE, which simply seeks for the best pulse for a set of different systems, SOMA SL, which first solves a sample of systems and then performs a regression over the sampled values -- for the sake of completeness we consider here both a linear and a non-linear approach -- and SOMA BP, where the network is trained directly on the average infidelity of an ensemble of systems with back propagation. \\
	In the case of Robust GRAPE, we sample $L$ systems and run the optimization with random restart, i.e., we run the optimization $N_{\text{guesses}}= 5$ times with different conditions and then choose the pulse with the smallest average infidelity.\\
	For SOMA SL, we sample up to $N = 10000$ points within the parameter space. For each one of these, we optimize the corresponding QOC problem with GRAPE -- this can be performed with any proper optimal quantum control method -- and then train the model to map the corresponding parameter to the optimal pulses. \\
	As for SOMA BP, we sample $L$ systems and run the neural-network training with random restart. The number of total samples required by the regression is usually larger than the one needed by the other two methods, the corresponding single-system optimization is nonetheless much faster and can be run in parallel. Therefore we set $N = 10000$ for each simulation, whereas for the sake of comparison and due to the similarity of the other two methods always use the same number of samples (either $L=500$ for the single qubit case or $L=100$ for the two-qubit case). Both SOMA methods employ two-layered neural networks with 256 neurons per layer for the single qubit system and 500 neurons per layer for the two-qubit system. The linear model performs multi-linear regression on the same data used by SOMA SL. \\
	For the random restart of SOMA methods, we run the optimization $N_{\text{guesses}} = 5$ times with different initial conditions and then test the quality of the predictions on a test set. Afterwards we choose the model with the lowest average test infidelity. 
	For both algorithms, we use the L-BFGS-B algorithm.
    Since the training of SOMA SL, which uses a MSE loss, is much faster than SOMA BP, we do not limit its maximal number of iterations. For SOMA BP, however, this is limited to 6000 for the single qubit gates and to 7000 for the two-qubit gates. Here, the input parameters for both SOMA SL and SOMA BP should be re-scaled to lie e.g., within the range [0,1]. For SOMA BP, this can be avoided in certain cases, as long as the size of the pulses remains large enough. For details about the performance of the algorithms, see Tab.~\ref{tab:alg_features}.\\
	For each system, we evaluate the performance of the neural-network pulses on the entire parameter space. In order to visualize the performance of the algorithms considered in a one-dimensional plot, we consider the normalized parameter space, where all parameters axes are re-scaled to the interval [0,1].
	\begin{align}
		f: \mathbb{R}^D \mapsto [0, 1]^D, \vec{\lambda} \longrightarrow \frac{\vec{\lambda} - \vec{\lambda}_{\text{min}}}{\abs{\vec{\lambda}_{\text{max}} - \vec{\lambda}_{\text{min}}}}
 	\end{align}
	By then considering $D$-dimensional spheres of radius $r \in [0,1]^D$, we sample $N_s = 1000$ systems on the surfaces of such spheres and compute the average infidelity over these systems. By doing so, we can evaluate the performance over the parameter space, thereby ensuring that our methods are effective for every combination of problem parameter values. The result is then averaged over $N_s$ samples, i.e., we plot the mean and its standard deviation. The latter is pictured as a shady region.  \\
	We also want to consider how the standard deviation of the average infidelity behaves when the different algorithms are tested against a batch of quantum systems sampled according to the parameter space. In particular, the standard deviation of the infidelity for $N_s$ test systems is bounded by the average performance of the algorithms:
	\begin{align}
	    \sigma(\bm{w})^2 = \frac{1}{L} \sum_{i=1}^{N_s} F(\bm{w}, \vec{\lambda}_i)^2 - F_{\text{test}}^2 \leq 1 - F_{\text{test}}^2
	\end{align}
	where $F_{\text{test}}$ is the average fidelity of the trained pulses on the $N_s$ test systems. For SOMA BP, assuming that over-fitting is negligible, we have $\sigma(\bm{w})^2 \leq 1 - (1-\mathcal{L}(\bm{w}))^2$.
	which means that the fidelity of the algorithms considered is guaranteed not to drop significantly below the average performance showed in Fig.~\ref{fig:4plots_1}, Fig.~\ref{fig:4plots_2} and Fig.~\ref{fig:CR_plots_1}.
	
	\red{\section{Fidelity of discrete gates and their performance}
	\label{appx:compilation}
	In this section we briefly illustrate how analog gates are expected to outperform equivalent circuit decomposition with sequences of traditional discrete gates. In particular, we first consider the standard universal set of quantum gates $S = \{\text{CNOT}, T, S, H \}$ used in Fig.~\ref{fig:SOMA}. We searched for decompositions  to determine the correct compilation sequence for the aforementioned gates. In particular, we consider the CR gate defined in Eq.~\eqref{eq:cr_theta} for an angle of $\theta = \frac{\pi}{4}$. The decomposition is given by the following circuit:
	
	\begin{center}
	    $\text{CR}(\frac{\pi}{4})$ =
        \begin{quantikz}
        & \ctrl{1} & \qw & \qw & \gate{S} & \qw \\
        & \targ{} & \gate{H} & \gate{S} & \gate{H} & \qw
        \end{quantikz}
	\end{center}

     This decomposition consists of 1 entangling gate (the CNOT) and 4 local unitary operations and the representation is exact. Note that CNOT has to be implemented efficiently on the chosen quantum computing platform, which probably anyway requires quantum control. An entangling gate in the context of superconducting circuits usually requires $T \sim 150 \ \si{ns}$ \cite{Kandala2021}. This is a little bit slower than the CR gate presented here but shows the equivalency of these gate sets. Nonethless, SOMA also allows learning both the CR and CNOT as part of the same continuous gate set, thereby saving the additional cost of single qubit gates (which take typically at least $T \sim 15 \ \si{ns}$ each).
     
      	\begin{table}[ht!]
	\begin{tabular}{|c||c|c|c|c|c|}
         \hline
         & \multicolumn{4}{|c|}{Number of gates} & \\
         \hline
         Angle $\theta$ & $N_{\text{CNOT}}$ & $N_{T}$ & $N_{S}$ & $N_{H}$ & highest fidelity \\
         \hline
         $\frac{\pi}{\sqrt{2}}$ &  1  &  2 & 4 & 1 & 0.9819   \\
         $\frac{\pi}{\sqrt{3}}$ &  2  &  3 & 1  & 2 & 0.9778      \\
         $\frac{\pi}{\sqrt{5}}$ &  0  &  2 & 3 & 2 & 0.9728 \\
         $\frac{\pi}{\sqrt{7}}$ &  2  &  3 & 0 & 2 & 0.9999\\
         \hline
    \end{tabular}
    \caption{Results of searching for CR gate decompositions for four different angles: $\frac{\pi}{\sqrt{2}}, \frac{\pi}{\sqrt{3}},\frac{\pi}{\sqrt{5}},\frac{\pi}{\sqrt{7}}$, obtained with the help of exhaustive search (up to 10 circuit layers) and stochastic descent (up to 20 circuit layers). The universal gate set is $S$. We observe here that while for $\frac{\pi}{\sqrt{7}}$ we can find a suitable decomposition with high fidelity, this is not the case for the remaining angle values. Moreover, two of the best decompositions contain two CNOT gates, which again lead us to much longer gate evolution time.}
    \label{tab:discrete_gates}
    \end{table}
    
     Where the advantage really shows up is for a gate angle that does not belong to the same entanglement class as CNOT.
     It is clear here, how continuously parametrized gates and consequently SOMA can be beneficial to quantum compilation.
     A CR gate with an angle $\theta = \frac{\pi}{8}$ also requires 5 gates to be implemented, as shown by the following circuit,
     \begin{center}
	    $\text{CR}(\frac{\pi}{8})$ =
        \begin{quantikz}
        & \qw & \ctrl{1} & \qw & \ctrl{1} & \qw \\
        & \gate{H} & \targ{} & \gate{S} & \targ{} &\gate{H}
        \end{quantikz}
	\end{center}
     but its circuit contains two CNOTs.  Since the CR($\frac{\pi}{8}$) can take roughly half the time of a CR($\frac{\pi}{4}$), this implies the CNOT-based circuit can be as much as 7 times slower than the continuously parametrized gate.

    \begin{figure}[ht!]
        \centering
        \hspace{-0.25cm}
        \includegraphics[width=8 cm]{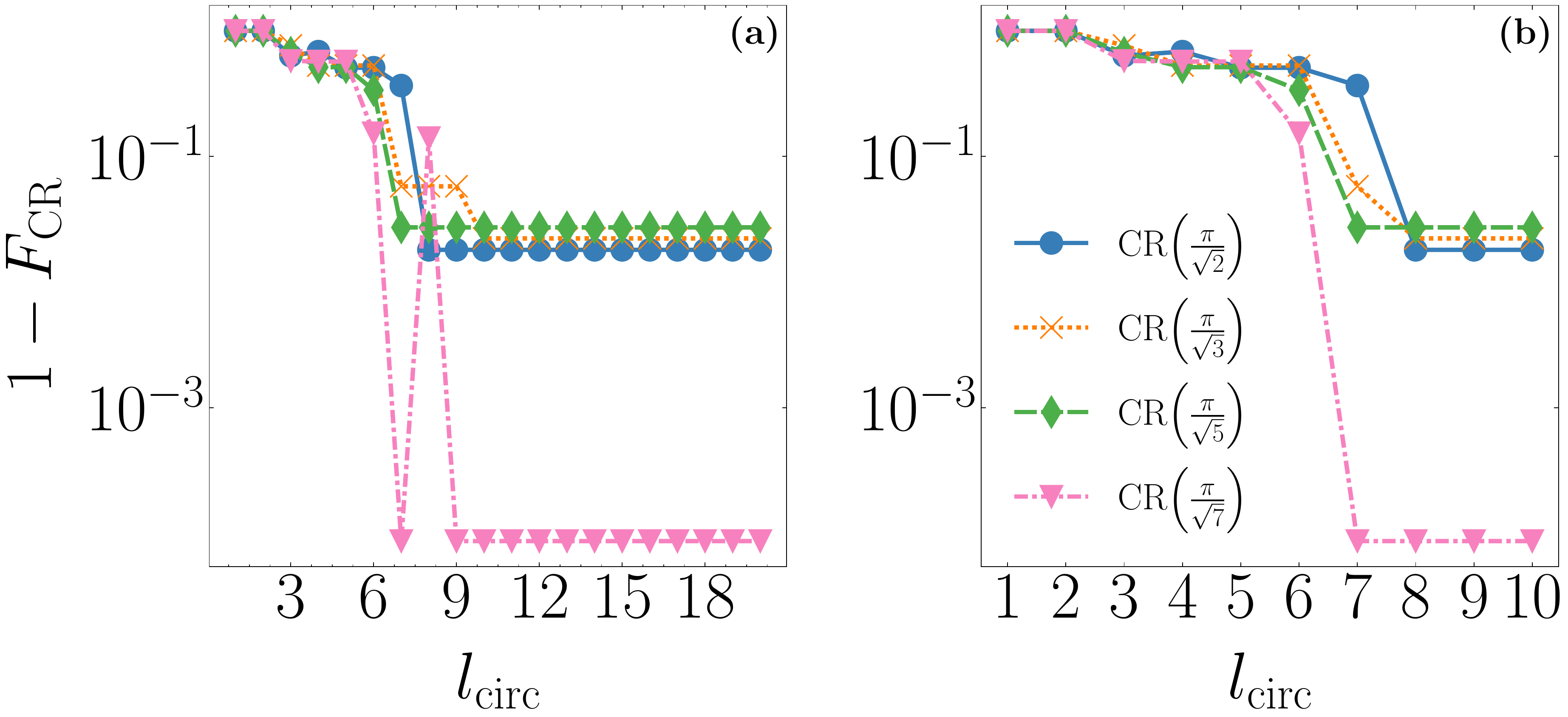}
        \caption{Infidelities reached by applying stochastic descent search (a) and brute-force (exhaustive) search (b) on the CR the gates defined by the angles in Tab.~\ref{tab:discrete_gates} in order to decompose them in discrete circuits. As we can observe, both algorithms provide us with the same results. However, they cannot find a circuit representation with $F>0.98$ for the first three gates, thus indicating that circuits decomposing these gates with a higher fidelity are longer, more error-prone, and harder to discover.}
        \label{fig:circ_search}
    \end{figure}
    
	For other angles, it becomes increasingly difficult to find good circuits, with only partial approximations being possible at reasonably short depth.  In order to test the quality of the approximation, we search for optimal discrete circuits representing a quantum circuit according to a given parametrization. We use exhaustive search \cite{Nievergelt2000} of all possible circuits (up to circuit depth 10) and stochastic descent, a special type of structured random search  -- see RL approach in \cite{Bukov2018} -- (up to circuit depth 20) to search for optimal decompositions of discrete gates and try to reproduce the chosen circuit with increasing number of unitaries. The results are given in Tab.~\ref{tab:discrete_gates} and Fig.~\ref{fig:circ_search} for the CR gate with angles $\frac{\pi}{\sqrt{2}}, \frac{\pi}{\sqrt{3}},\frac{\pi}{\sqrt{5}},\frac{\pi}{\sqrt{7}}$. We see that although the fidelity of the discrete gates increases with the size of the quantum circuit, in three cases it cannot reach the value of F=0.99 for circuits of depth smaller than 20. In the case of $\theta = \frac{\pi}{\sqrt{7}}$, a valid decomposition with fidelity F=0.9999 is found. The search is performed exhaustively for $l_{\text{circ}} < 10$ and then using stochastic descent for $l_{\text{circ}} > 10$. Moreover, the decomposition of $\text{CR}(\frac{\pi}{\sqrt{7}})$ contains 2 CNOT gates, which again implies a slowdown of the gate execution time.
	Other examples for gates with superior analog performance can be seen, e.g., Ref.~\cite{foxen2020demonstrating}, with superior performance especially expected for variational circuits.
}
\end{document}